%% file: main.tex
\documentclass[12pt]{article}
\usepackage[utf8]{inputenc}

\usepackage{amsmath}
\usepackage{amsthm}
\usepackage{amsfonts}
\usepackage[table]{xcolor}    
\usepackage{makecell}

\newcommand{\bem}[2]{\textcolor{#1}{{#2}}}
\definecolor{midnight}  {rgb}{0,0,.5} 

\usepackage{times}
\usepackage{epsfig}
\usepackage{graphicx}
\usepackage{amsmath}
\usepackage{tikz}
\usepackage{mdframed}

\usepackage{amssymb}
\usepackage[caption=false]{subfig}
\usepackage{bm}
\usepackage[breaklinks=true,bookmarks=false]{hyperref}
\usepackage[normalem]{ulem}
\usepackage{multirow}
\usepackage{tabu}
\usepackage{enumitem}
\usepackage{adjustbox}
\usepackage{cite}
\usepackage[margin=1in]{geometry}
\usepackage{cleveref}

\setlist[itemize]{leftmargin=*}

\input{macros-new}

\newcommand{\R}{\mathbb{R}}
\newcommand{\E}{\mathbb{E}}
\newcommand{\divg}{\text{div}}
\DeclareMathOperator*{\argmax}{arg\,max}

\title{
\LARGE\bf Deep Learning Techniques \\
for Inverse Problems in Imaging}

\author{
Gregory Ongie\footnote{
Department of Statistics, University of Chicago, Chicago, IL.},~
Ajil Jalal\footnote{Department of Electrical and Computer Engineering, University of Texas at Austin, Austin, TX.},~
Christopher A.\ Metzler\footnote{Department of Electrical Engineering, Stanford University, Stanford, CA.}\\
Richard G.\ Baraniuk\footnote{Department of Electrical and Computer Engineering, Rice University, Houston, TX.},~
Alexandros G. Dimakis\footnote{Department of Electrical and Computer Engineering, University of Texas at Austin, Austin, TX.},~
Rebecca Willett\footnote{
Departments of Statistics and Computer Science, University of Chicago, Chicago, IL.} 
}

\date{April 2020}

\begin{document}

\maketitle

\begin{abstract}
Recent work in machine learning shows that deep neural networks can be used to solve a wide variety of inverse problems arising in computational imaging. We explore the central prevailing themes of this emerging area and present a taxonomy that can be used to categorize different problems and reconstruction methods. Our taxonomy is organized along two central axes: (1) whether or not a forward model is known and to what extent it is used in training and testing, and (2) whether or not the learning is supervised or unsupervised, i.e., whether or not the training relies on access to matched ground truth image and measurement pairs. We also discuss the tradeoffs associated with these different reconstruction approaches, caveats and common failure modes, plus open problems and avenues for future work.
\end{abstract}

\section{Introduction}

This paper concerns \textit{inverse problems}, i.e., reconstructing an unknown signal, image, or multi-dimensional volume from observations. The observations are obtained from the unknown data by a \textit{forward process}, which is typically non-invertible.
Numerous imaging tasks fit under this framework, including image deblurring, deconvolution, inpainting, compressed sensing,  superresolution, and many more.  
These forward processes are ill-posed, so reconstructing a \textit{unique} solution that fits the observations is difficult or impossible without some prior knowledge about the data. 
Traditional methods minimize a cost function that consists of a data-fit term, which measures how well the reconstructed image matches the observations, and a regularizer, which reflects prior knowledge and promotes images with desirable properties like smoothness. 
Deep learning techniques are currently transforming image reconstruction methods and impact applications ranging from geophysical, scientific and medical imaging. 
We provide an overview of this rapidly evolving landscape. 

To be more precise, we consider inverse problems in which an unknown
$n$-pixel image (in vectorized form) $\bxstar \in \reals^n$ (or $\mathbb{C}^n$) is observed via $m$ noisy measurements 
$\by\in\mathbb{R}^m$ (or $\mathbb{C}^m$) according to the model
\[
\by=\mathcal{A}(\bxstar)+\bm\varepsilon,
\]
where $\mathcal{A}$ is the (possibly nonlinear) forward measurement operator and $\bm\varepsilon$ represents a vector of noise. The goal is to recover $\bxstar$ from $\by$. More generally, we can consider non-additive noise models of the form
\[
\by=\mathcal{N}(\mathcal{A}(\bxstar)),
\]
where $\mathcal{N}(\cdot)$ samples from a noisy distribution.
Without loss of generality, we assume that $\by, \bxstar, \mathcal{A},$ are real-valued, since most techniques presented in this paper can be generalized to complex-valued images/measurements by concatenating real and imaginary parts. 

This general model is used throughout computational imaging\cite{barrett2013foundations}, from basic restoration tasks like deblurring, super-resolution, and image inpainting \cite{dip}, to a wide variety of tomographic imaging applications, including magnetic resonance imaging \cite{fessler2010model}, X-ray computed tomography \cite{elbakri2002statistical}, and radar imaging \cite{blahut2004theory}. 
The task of estimating $\bxstar$ from $\by$ is often referred to as {\em image reconstruction}. Classical image reconstruction methods assume some prior knowledge about $\bxstar$ such as smoothness \cite{tychonoff1977solution}, sparsity in some dictionary or basis \cite{figueiredo2005bound,mairal2009online,yu2012solving,1710377}, or other geometric properties \cite{rudin1992nonlinear,willett2003platelets,danielyan2012bm3d,marais2017proximal, tamir2020computational}. 
Reconstruction amounts to finding an $\bxhat$ that is both
a good fit to the observations $\by$ and also 
likely given the prior knowledge. A {\em regularization function} $r(\bx)$ measures the lack of conformity of $\bx$ to a prior model and $\bxhat$ is selected so that $r(\bxhat)$ is as small as possible while still fitting the observed data. 

Recent work in machine learning has demonstrated that deep neural networks can leverage large collections of training data to directly compute regularized reconstructions across a host of computational imaging tasks~\cite{mousavi2015deep,kulkarni2016reconnet,dong2016image,sun2016deep,mousavi2017learning,zhang2017beyond,rick2017one,mousavi2017deepcodec,hammernik2018learning}. 
In parallel, work on unsupervised methods showed how deep generative models can regularize by constraining the reconstructed image $\bxhat$ to remain on a learned manifold~\cite{bora2017compressed}.
We explore the central prevailing themes of this emerging area and present a taxonomy that can be used to categorize different problems and reconstruction methods. 
We also discuss the tradeoffs associated with different reconstruction approaches and describe avenues for future work.

\section{Opportunities and Recent Progress}

In the last five years deep learning has demonstrated enormous potential for solving a host of imaging inverse problems; see, e.g., \cite{lucas2018using}. However, a fundamental understanding of the applicability of deep learning methods and their limitations remains in its infancy. This creates opportunity for additional research, careful scientific evaluation, and foundational understanding.

\begin{figure}[ht!]
    \centering
    \includegraphics[width=0.8\columnwidth]{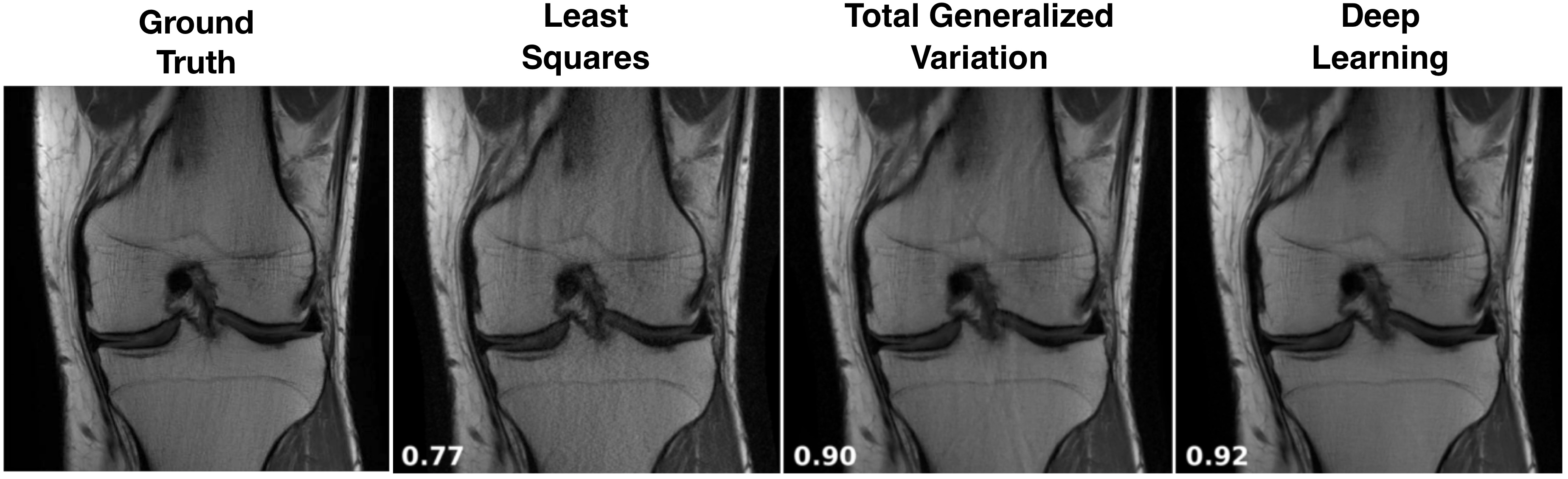}
    \caption{\textbf{Undersampled MRI reconstruction.}  Undersampled multi-coil MRI reconstruction using least-squares reconstruction, total generalized variation regularization, and a deep learning approach. SSIM quality measures are displayed in white. This example highlights the potential of learned methods of image reconstruction over more conventional techniques. Figures adapted from~\cite{knoll2020deep}.}
    \label{fig:mri}
\end{figure}

\paragraph{Medical imaging.} Reconstructing images from projective measurements arises in MRI, CT, PET, SPECT, and many other modalities. Classical methods perform well but can be computationally demanding. Recent work on using training data to improve the reconstruction process can lead to better image quality and orders of magnitude faster reconstructions than classical iterative methods. The potential gains are illustrated in Figure~\ref{fig:mri} and an excellent overview was recently published \cite{knoll2020deep}.
GE's ``TrueFidelity'' deep learning image reconstruction for CT imaging \cite{trueFidelity}
has FDA approval as of April 2019 \cite{trueFidelityFDA}.
Nevertheless, numerous open questions remain, as described in Section~\ref{sec:caveats}.

\begin{figure}[ht!]
    \centering
    \includegraphics[width=0.8\columnwidth]{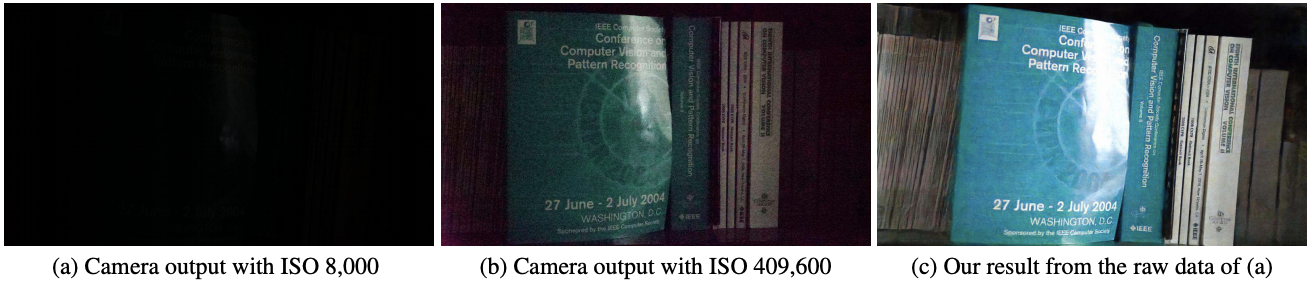}
    \caption{\textbf{Low-light imaging.} A network takes an underexposed image (left) and denoises and white balances it to produce a clean image (right) that does not exhibit the color bias associated with extreme ISO images (middle). Figures adapted from~\cite{chen2018learning}.}
    \label{fig:seeindark}
\end{figure}

\begin{figure}[!ht]
    \centering
    \includegraphics[width=0.9\columnwidth]{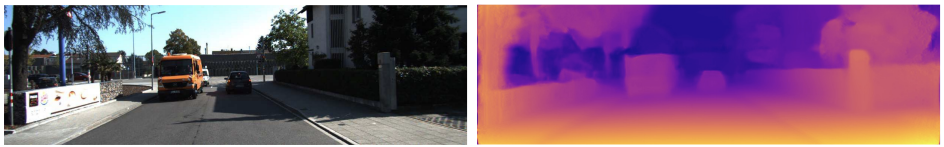}
    \caption{\textbf{Monocular depth estimation.} A network takes an image (left) and produces a depth map (right). Figures adapted from~\cite{godard2017unsupervised}.}
    \label{fig:monocular}
\end{figure}

\paragraph{Computational photography.}
The goal of computational photography is to create visually appealing images that are reasonably faithful to the scenes they represent. These conditions make deep learning an excellent candidate for computational photography reconstruction problems.
For example, deep learning enables exceptional low-light imaging~\cite{chen2018learning}, as shown in Figure \ref{fig:seeindark}. Deep learning also enables estimating the depths of different objects in a scence from a photograph \cite{godard2017unsupervised}, as illustrated in Figure~\ref{fig:monocular}.
Presently, deep learning is used to perform white balancing in the production version of Google's latest smart phone imaging systems~\cite{liba2019handheld,barron2017fast}.

\paragraph{Computational microscopy.}
With the growing popularity of computational techniques like ptychography, solving a reconstruction problem has become an integral part of microscopy. Accordingly, there has been a surge of interest in applying deep learning to microscopy, producing new techniques to both reconstruct images and design a microscope's illumination patterns and optical elements~\cite{kamilov2015learning,rivenson2017deep,nehme2018deep,pinkard2019deep,horstmeyer2017convolutional,hershko2019multicolor,kellman2019data}. 

\begin{figure}[ht!]
    \centering
    \includegraphics[width=0.8\columnwidth]{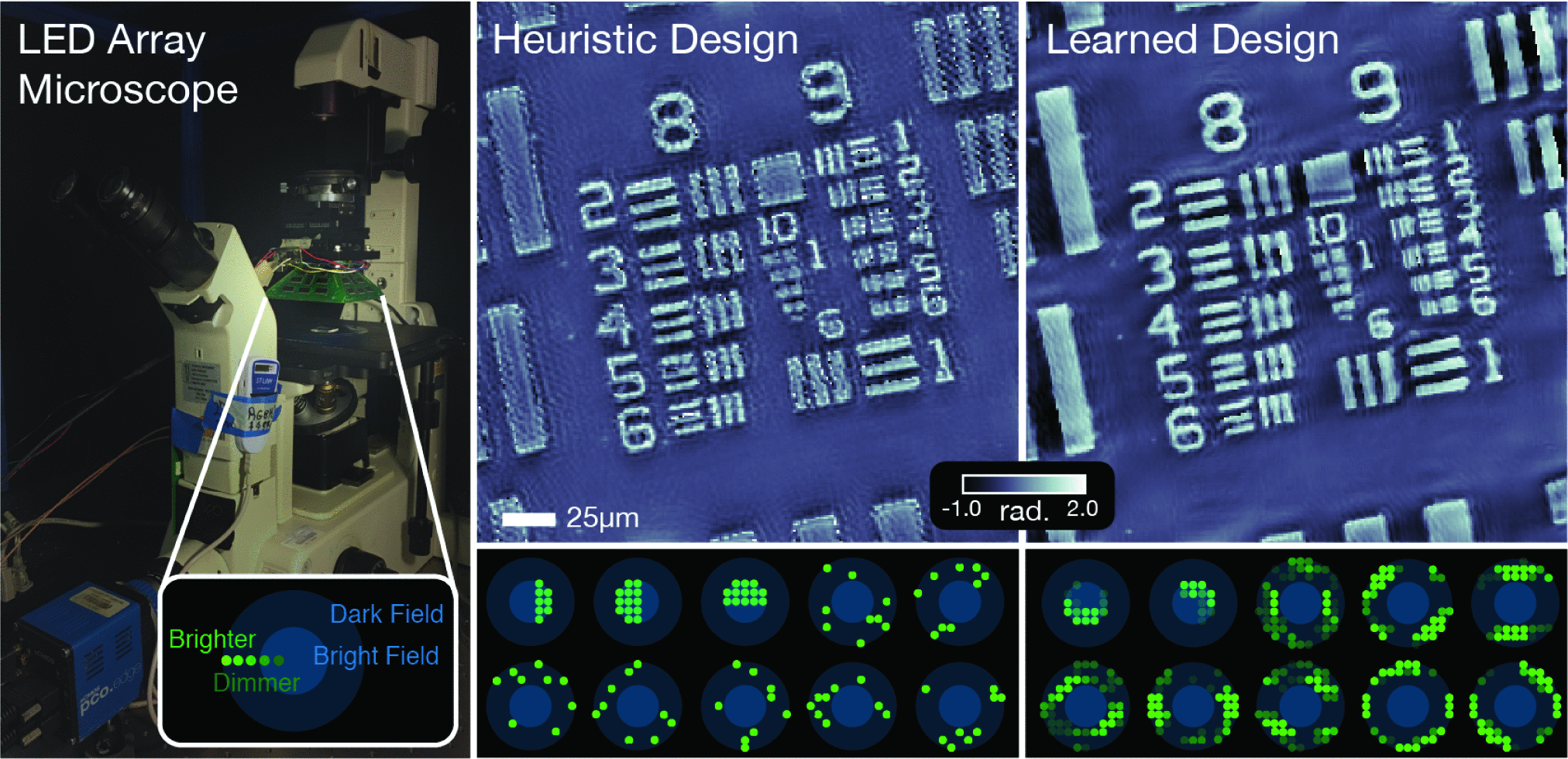}
    \caption{\textbf{Microscopy with learned illumination patterns.} Differentiable optical models and reconstruction algorithms are used to design illumination patterns (bottom row) for Fourier ptychographic microscopy. Figures adapted from~\cite{kellman2019data}.}
    \label{fig:microscopy}
\end{figure}

 \paragraph{Geophysical imaging.}

Seismic inversion and imaging involves reconstructing the 
Earth’s interior by modeling the physical propagation of seismic waves.
The comparison of simulated synthetic measurements to actual acoustic recordings of reflected waves can be used to tune the models of these ill-posed inverse problems. 
Deep learning techniques have been recently proposed to tackle these problems~\cite{araya2018deep,hansen2017efficient}, including methods that rely on generative models~\cite{mosser2020stochastic} constrained by partial differential equations; see also~\cite{yang2020physics}.

\begin{figure}[ht!]
    \centering
    \includegraphics[height=0.3\columnwidth]{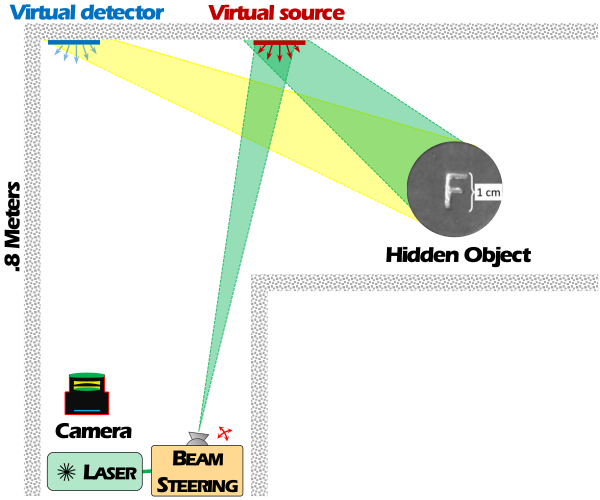}
    \quad\includegraphics[height=0.3\columnwidth]{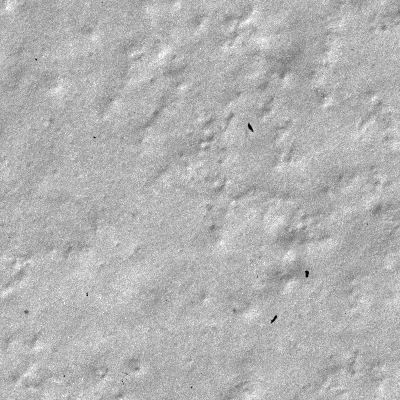}
    \quad\includegraphics[width=0.1\columnwidth]{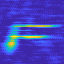}
    \caption{\textbf{Imaging around corners.} In deep-inverse correlography, deep learning is used to reconstruct a hidden object (right) from a series of speckle images (middle). Figures adapted from~\cite{metzler2020deep}.}
    \label{fig:NLOS}
\end{figure}

\paragraph{Other computational imaging applications.} 
While still in the development stages, deep learning has shown immense promise in many other challenging computational inverse problems, including lensless imaging~\cite{sinha2017lensless}, holography~\cite{rivenson2018phase,wu2018extended}, ghost imaging~\cite{lyu2017deep}, imaging through scattering media~\cite{borhani2018learning,li2018deep,sun2018efficient}, and non-line-of-sight imaging~\cite{metzler2020deep}, which is illustrated in Figure \ref{fig:NLOS}. See~\cite{barbastathis2019use} for a recent optics-focused review article.

\section{Anatomy of an Inverse Problem}

Recall that we are interested in recovering a vectorized image $\bx\in \mathbb{R}^n$  from measurements $\by\in\mathbb{R}^m$ of the form
$\by=\mathcal{A}(\bx)+\bm\varepsilon$,
where $\mathcal{A}$ is the (possibly nonlinear) forward measurement operator and $\bm\varepsilon$ represents noise. Alternatively, the system can be represented by
$\by=\mathcal{N}(\mathcal{A}(\bx))$, where $\mathcal{N}(\cdot)$ samples from a noisy distribution. 

If the distribution of the noise is known, solving a maximum likelihood (ML) estimation problem can recover $\bx$:
\[
\hat{\bx}_{\rm ML} = \argmax_{\bx} p(\by|\bx) = \argmin_{\bx} -\log p(\by|\bx),
\]
where $p(\by|\bx)$ is the likelihood of observing $\by$ if $\bx$ were the true underlying image. (Knowledge of $\cA$ or integration over a distribution of possible $\cA$'s is implied in this formulation.) The maximum likelihood approach has some significant drawbacks, including potentially non-unique solutions (\eg when $\cA$ is a linear operator with rank less than $n$) or high sensitivity to noise (\eg when the spectrum of $\cA$ is not bounded below; in the case where $\cA$ is the linear operator $A$, this corresponds to some eigenvalues of $A^\top A$ being small). 

In some settings, one might have prior knowledge about which $\bx$ are more likely; for instance, we might expect $\bx$ to be smooth, or be smooth away from edges and boundaries. Such knowledge can be codified into a prior distribution for $\bx$, leading to a {\em maximum a posteriori} (MAP) estimate
\[
\hat{\bx}_{\text{MAP}} = \argmax_{\bx} p(\bx|\by) = \argmax_{\bx} p(\by|\bx)p(\bx) =\argmin_{\bx} -\ln p(\by|\bx) - \ln p(\bx).
\]
For the special case of additive white Gaussian noise, the MAP formulation leads to
\begin{align}\label{eqn:MAP}
\argmin_{\bx}\tfrac{1}{2}\|\mathcal{A}(\bx)-\by\|^2_2 + r(\bx),
\end{align}
where $r(\bx)$ is proportional to the negative log-prior of $\bx$. 
Examples of this framework include Tikhonov regularization \cite{tikhonov1943stability}, sparsity regularization in some basis or frame \cite{wright2009sparse,tropp2010computational}, and total variation regularization \cite{rudin1992nonlinear,wang2008new}.
In some settings, MAP estimation with underdetermined $\mathcal{A}(\cdot)$ can be considered an algorithmic procedure for choosing, among the infinitely many values of $\bx$ that satisfy $y=\mathcal{A}(\bx)$, the one that is most likely under the prior.

While in principle MAP estimation can be used to solve most image reconstruction problems, difficulties arise when (1) the statistics of the noise are not known, (2) the distribution of the signal is not known or the log-likelihood does not have a closed form, or (3) the forward operator is not known or only partially known. 
In the last five years, machine learning has provided machinery to (partially) overcome many of these issues. 
Variations on the aforementioned inverse problem appear in a range of imaging settings. We highlight a few prominent examples in Table~\ref{tab:applications}.
\input{applications.tex}

\subsection{Supervised vs.\ Unsupervised Inversion}

We start by explaining a central dichotomy in the literature and in our proposed taxonomy of approaches to inverse problems. 
The first (and most well-known) family of 
deep learning inversion methods use what we call \textbf{supervised} inversions. 
The central idea is to create a \textit{matched} dataset of ground truth images $\bx$ and corresponding measurements $\by$, which can be done by simulating (or physically implementing) the forward operator on clean data, i.e. measure them. Subsequently, one can train a network that takes in measurements $\by$ and reconstructs the image $\bx$, i.e. learns an inverse mapping. Such supervised methods typically perform very well, but are sensitive to changes or uncertainty to the forward operator $\mathcal{A}$. In addition, a new network needs to be trained every time the measurement process changes. 

The second family of techniques we cover are \textbf{unsupervised}, i.e. do not rely on a matched dataset of images $\bx$ and measurements $\by$. 
In our taxonomy we separate unsupervised methods into three different kinds:
(1) methods which use unpaired ground truth images $\bx$'s and measurements $\by$'s, (2) methods which leverage ground truth images $\bx$'s only, and (3) methods which use only measurements $\by$'s. 

\subsection{Background on Deep Generative Models}

A central challenge in the foundations of learning is succinctly modeling high-dimensional distributions in a way that permits efficient learning and inference. 
Simply put, the difficulty is that representing a general joint probability distribution over $n$ variables, even for binary random variables, requires $2^n-1$ parameters. Therefore, 
we must postulate some type of \textit{structure} on the data to overcome its worst-case complexity. 

Previous efforts towards this goal have been at the heart of substantial breakthroughs.  For example, in compressed sensing and high-dimensional statistics, the notion of sparsity and low-rank are key structural assumptions in many prior works. 
Sparsity (e.g., in Discrete Cosine Transform (DCT) or wavelet domains) plays \textit{the central role} in most lossy compression standards like JPEG~\cite{wallace1992jpeg}, JPEG-2000~\cite{marcellin2000overview} and MPEG~\cite{manjunath2002introduction}.
Another successful example is graphical models: 
in this case a high dimensional distribution becomes tractable through factorization, which is equivalent to conditional independence. For Bayesian networks and undirected graphical models there is a rich theory for both learning and inference including precise conditions under which structure learning can be achieved efficiently. 

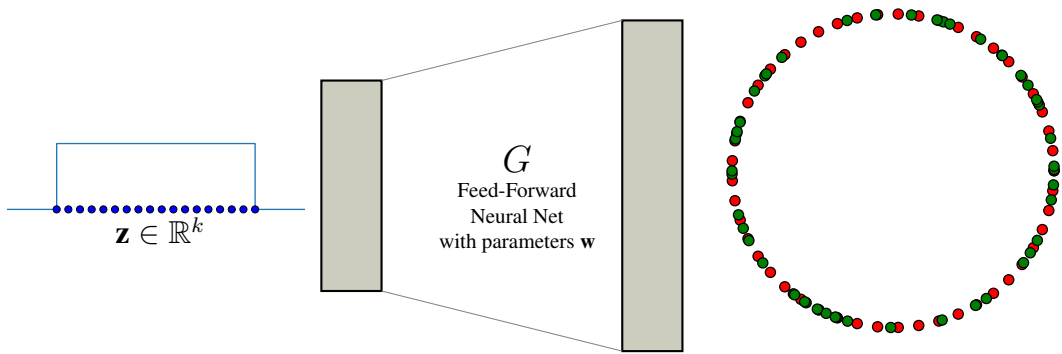
\begin{figure}[ht!]
\rowcolors{2}{white}{white}
\begin{center}
	\vspace{0.5cm}
 \begin{tikzpicture}[xscale=4, yscale=4]	 
\definecolor{mycolor}{rgb}{0.122, 0.435, 0.698}
\definecolor{myBurntOrange}{rgb}{0.75, 0.33, 0.0}
\definecolor{myBlue}{rgb}{0.2, 0.25, 0.28}  
\definecolor{myGray}{rgb}{0.8, 0.8, 0.75}  
\newmdenv[innerlinewidth=0.5pt, roundcorner=4pt,linecolor=mycolor,innerleftmargin=6pt,backgroundcolor=yellow!40,innerrightmargin=6pt,innertopmargin=6pt,innerbottommargin=6pt]{mybox}
\filldraw[fill=myGray, draw=black,thick] (-0.5,0.3) rectangle (-0.3,1);

\filldraw[fill=myGray, draw=black,thick] (0.5,0.1) rectangle (0.7,1.2);
\draw [help lines] (-0.3,0.3) -- (0.5,0.1);   
\draw [help lines] (-0.3,1) -- (0.5,1.2);   
\node[inner sep=0pt] at (-1.05,0.9)
    {\includegraphics[width=.25\textwidth]{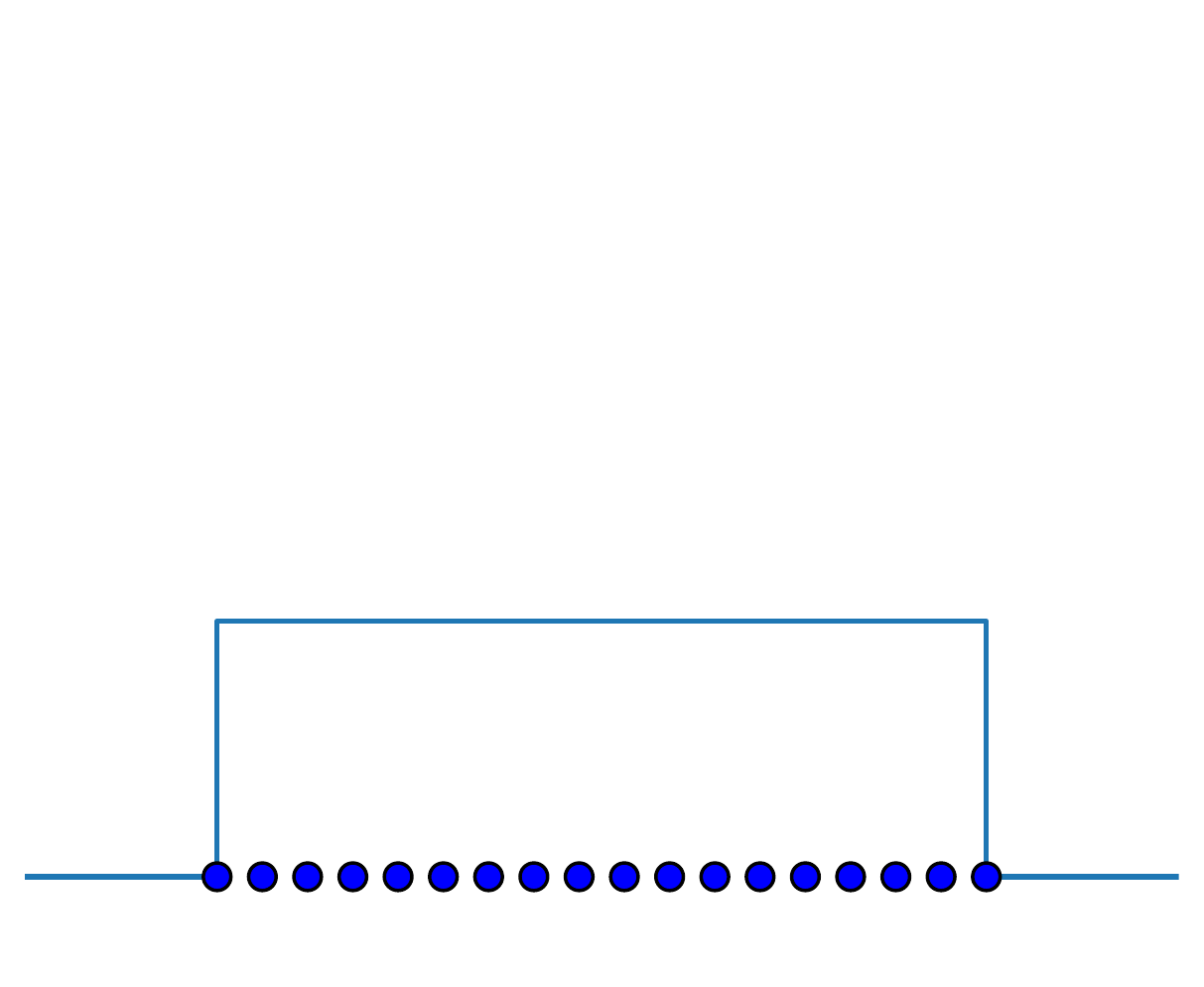}};
\node [left] at (-0.85, 0.5) {$\textbf{z} \in \mathbb{R}^k$ };   
 
\node[inner sep=0pt] at (1.40,0.70)
    {\includegraphics[width=.30\textwidth]{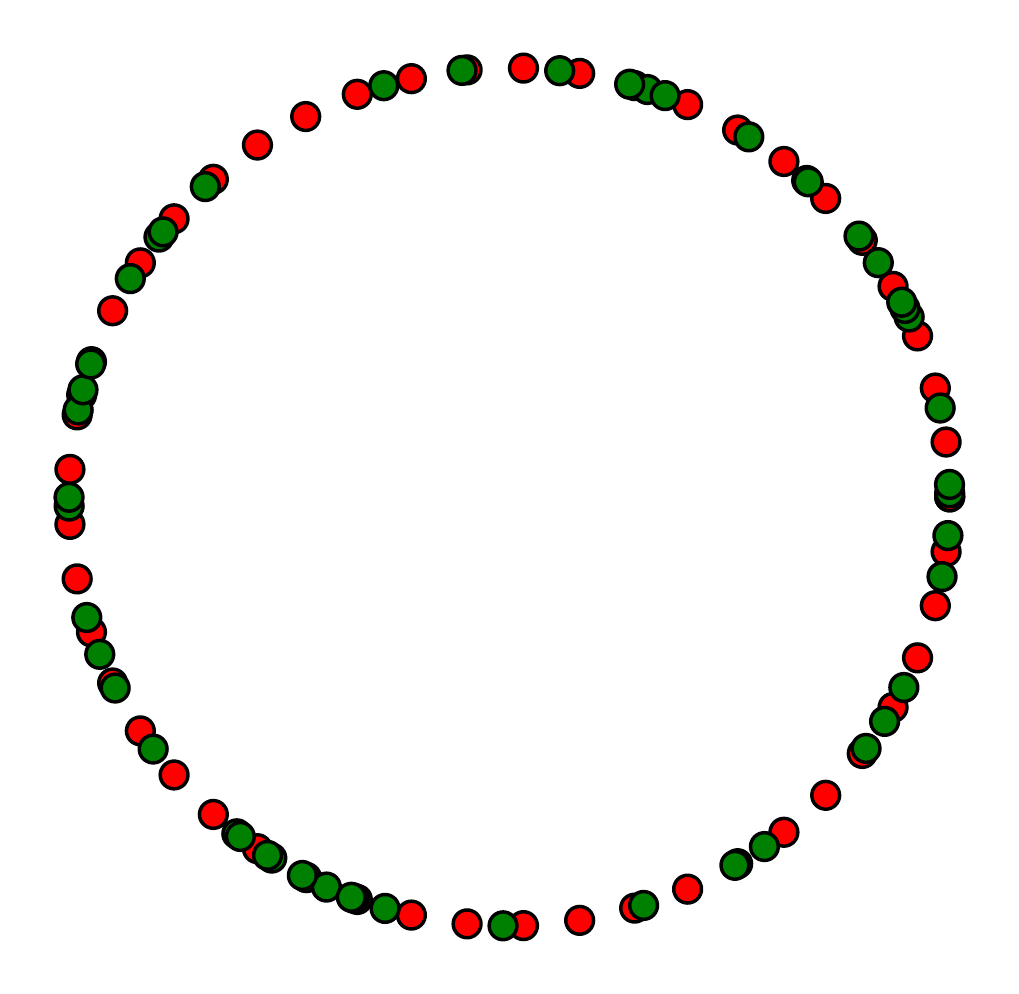}};

\node [left] at (0.5, 0.6) {\scriptsize 
\begin{tabular}{c} {\large $G$} \\ Feed-Forward \\ Neural Net   \\ with parameters $\textbf{w}$ \end{tabular}   
};
\vspace{0.2cm}
\end{tikzpicture}
\end{center}
    \caption{
A deep generative model is a function $G(\textbf{z})$ that takes a low-dimensional random vector $\textbf{z}$ in $\mathbb{R}^k$ and produces a high-dimensional sample 
$G(\textbf{z}) \in \mathbb{R}^n$.
In the example shown in the figure, the generator $G:\R\rightarrow\R^2$ learns to map low-dimensional samples $\textbf{z}$ (blue dots) drawn from a uniform distribution (blue line), such that the distribution of $G(\textbf{z})$ (green dots) resembles the distribution of training samples (red dots). 
While the output of this simple generative model lies in $\R^2$, modern deep generative models are capable of generating images with millions of pixels.
The low dimensional samples are typically drawn from a distribution which is easy to sample from, such as i.i.d.\ Gaussian or uniform distributions. 
Furthermore, the function is usually a convolutional neural network and is therefore continuous and differentiable almost everywhere. 
}
\label{figure:dgm}
\end{figure}  

Here, we are interested in a different way of modeling high-dimensional distributions: \textit{Deep Generative Models} (DGMs).
DGMs represent a complex distribution using a deterministic transformation applied to a simple ``seed'' distribution (e.g., independent Gaussian).
Formally, a DGM is described by a function
$G: \mathbb{R}^k \to \mathbb{R}^n$
parametrized by a deep neural network (typically convolutional) that is trained from actual data in an unsupervised way. Two primary DGM examples are Variational Auto-Encoders (VAEs)~\cite{kingma2013auto} and Generative Adversarial Networks (GANs)~\cite{goodfellow2014generative}.
DGMs are demonstrating unprecedented visual results for image generation, but many central theoretical questions about them remain poorly understood. We will discuss how DGMs can be used as priors for inverse problems~\cite{bora2017compressed,bora2018ambientgan,kamath2019lower}. 
Figure~\ref{figure:dgm} provides an illustrative example of a deep generative model that maps from $\R \rightarrow \R^2$.

\section{Taxonomy for Learning to Solve Inverse Problems}
\label{sec:taxonomy}
In recent years there has been an emerging body of literature on using training data to solve inverse problems in imaging. Some methods use the MAP formulation, seen in Equation~\eqref{eqn:MAP} as a starting point and attempt to learn the regularizer $r$ or some functional of $r$, while others attempt to directly learn a mapping from measurements $\by$ to images $\bxhat$. In this section, we describe a taxonomy for these approaches that facilitates an easier comparison among different methods and a better understanding of the tradeoffs among them. The overall taxonomy is shown in Table~\ref{tab:taxonomy} and details are described below. 

Computational imaging techniques often rely on a \emph{forward model} $\cA$, \ie a computational model of the physics underlying the measurement process.  
A key distinction between many types of learned inverse problem solvers is what is known when about the forward model. Options include:
\begin{squishlist}
\item $\cA$ is known from the beginning (\ie even during the training process). Examples include the discrete Radon and X-ray transforms in computed tomography, and the discrete Fourier transform in magnetic resonance imaging.
\item $\cA$ is not known during training, but after training may be used at test time (\ie during the reconstruction process). This framework is useful for training a general purpose model that can be ``plugged in'' to a variety of reconstruction tasks. 
\item $\cA$ is partially known. For instance, it might rely on calibration parameters that are unknown or difficult to estimate, as is the case in blind deconvolution problems encountered in optical imaging.
\item $\cA$ is never known or modeled. In this case, all information about $\cA$ is represented in the training data.
\end{squishlist}
Each of these settings requires different methods and analyses; we elaborate below. 
If an accurate forward model is known -- even partially -- then one might argue it should be used during training so that parameters are not wasted on ``learning the physics''. Indeed, several studies show that making effective use of forward models in training and testing dramatically reduces the sample complexity of learning-based image reconstruction. However, even if the forward model is known, it may be computationally prohibitive to apply. This becomes especially problematic in the training phase, where each backpropagation step may require multiple applications of the forward model or its adjoint. In this case, reconstruction architectures need to be carefully designed to reduce the number of applications the forward model or its adjoint.

\input{taxonomy_table}

\subsection{Forward Model Fully Known During Training and Testing}
\label{sec:Aknown}

When the forward model $\cA$ is fully known\footnote{Here we assume we know the ``true'' $\cA$ and not an approximation; issues related to only approximately knowing $\cA$ are discussed in Section \ref{sec:caveats}}, a wide variety of deep learning techniques can be employed to solve the inverse problem of interest. Here we will focus on the supervised setting where one has access to ground truth image/measurement pairs. We do not lose too much generality in our discussion by focusing on the supervised setting, since in unsupervised settings where one has access to ground truth images it is trivial to generate training pairs by applying the known forward model. However, the unsupervised setting where one only has access to (noisy) measurements requires novel techniques, which we address below. Finally, in Section~\ref{sec:open} we also discuss cases where one also has control over the design of $\cA$.

\subsubsection{Train from  $(\bx,\by)$ pairs (paired ground truth and measurements)}
\label{sec:1a}
The goal in a supervised setting is to estimate a \emph{reconstruction network} $f_\theta(\cdot)$ that maps measurements $\by$ to images $\bx$, where $\theta$ is a vector of parameters to be estimated from training data (e.g., neural network weights). Different deep learning approaches in the supervised setting can be thought of as different ways to parameterize the reconstruction network $f_\theta$. Specifically, $\cA$ itself (or mappings related to $\cA$ such as adjoints or derivatives) can be embedding into the architecture defining $f_\theta$. For simplicity, below we will assume $\cA$ is a linear operator, denoted by $A$, though many of the approaches we discuss extend naturally to nonlinear operators, as well.

\begin{figure}
    \centering
    \includegraphics[width=.5\columnwidth]{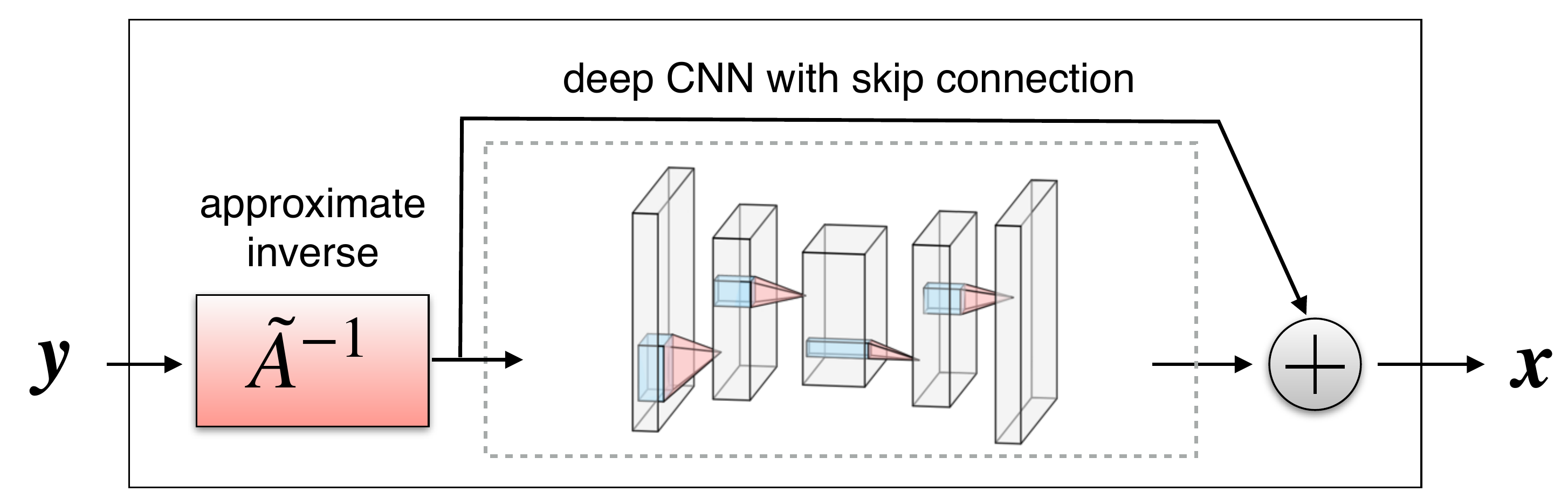}
    \caption{When an approximate inverse $\tilde A^{-1}$ of the forward model is known, a common approach in the supervised setting is to train a deep CNN to remove noise and artifacts from an initial reconstruction obtained by applying $\tilde A^{-1}$ to the measurements.}
    \label{fig:residual}
\end{figure}

One simple method of incoroporating knowledge of $A$ into the reconstruction network is by applying an approximate inverse of $A$, which we denote by $\widetilde{A}^{-1}$ (i.e., a matrix such that $\widetilde{A}^{-1}A\bx \approx \bx$ for all images $\bx$ of interest), to first map the measurements back to image domain and then train a neural network to remove artifacts from the resulting images. The specific choice of $\widetilde{A}^{-1}$ will depend on the particular inverse problem, but common choices include the adjoint $A^\T$ or psuedo-inverse $A^\dagger$, though one is not limited to these. For example, in super-resolution a common choice of $\widetilde{A}^{-1}$ is upsampling by bicubic interpolation \cite{dong2016image}; in CT reconstruction, a common choice of $\widetilde{A}^{-1}$ is filtered back projection \cite{mccann2017convolutional}.
This approach can be viewed as learning a reconstruction network whose first-layer weights are fixed and given by $\widetilde{A}^{-1}$. In this case, it is often beneficial to use a residual (or ``skip'') connections in the reconstruction network, since the output from the first layer is expected to be close to the output of the network. More precisely, this approach structures the reconstruction map $f_\theta$ as 
\begin{equation}    f_\theta(\by) = g_\theta(\widetilde{A}^{-1}\by) + \widetilde{A}^{-1}\by
\end{equation}
where $g_\theta$ is a trainable neural network depending on parameters $\theta$; see Figure \ref{fig:residual} for an illustration. In this case, the network $g_\theta$ is interpreted as predicting the residual between the approximate inverse and the reconstructed image. For example, in a super-resolution context, the network $g_\theta$ is predicts the missing high frequency content from a low-pass-filtered image. Networks with more complicated hierarchical skip connections are also commonly used, including the U-net \cite{ronneberger2015u} and architectures inspired by wavelet decompositions \cite{ye2018deep}.

\begin{figure}
    \centering
    \includegraphics[width=\columnwidth]{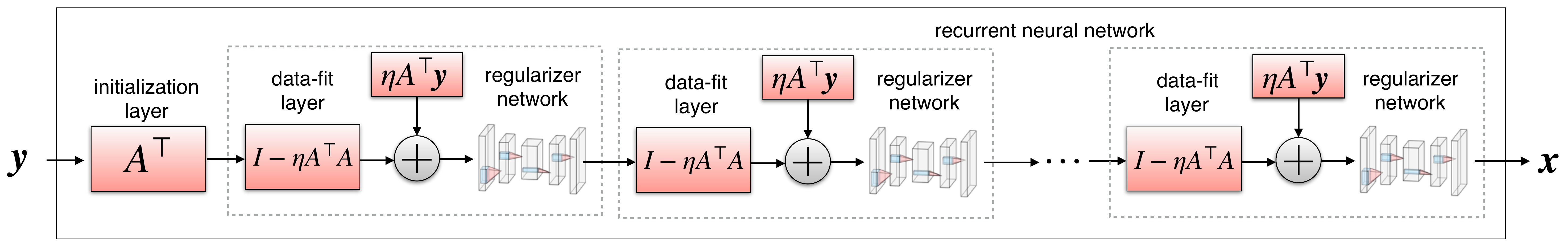}
    \caption{If the forward model $A$ and its adjoint $A^\top$ are known, then an iterative or unrolled network embeds $A$ and $A^\top$ in multiple layers of the network using a recurrent block artitecture. Here this approach is illustrated with an unrolling of the proximal gradient descent algorithm,  where the proximal map is replaced with a deep CNN.}
    \label{fig:unrolled}
\end{figure}

Inspired by iterative optimization approaches, \emph{unrolled} methods go further and incorporate $A$ into multiple layers of the reconstruction network, as illustrated in Figure~\ref{fig:unrolled}. 
To motivate this approach, consider the MAP formulation \eqref{eqn:MAP} where the regularizer $r(\cdot)$ (or, equivalently, the negative log-prior) is convex.
A commonly-used algorithm for optimizing \eqref{eqn:MAP} in this case is proximal gradient descent \cite{combettes2011proximal}, whose iterations have the form:
\begin{equation}\label{eq:gdn}
\bx^{(k+1)} = P\left( \bx^{(k)} -\eta A^\T(A \bx^{(k)}-\by) \right)
\end{equation}
where $P(\bz) := \argmin_{\bx} \{ \frac{1}{2}\|\bx-\bz\|^2 + r(\bx)\}$ denotes the proximal operator corresponding to the regularizer $r(\cdot)$, and $\eta$ is a step-size parameter. Suppose that we take as our reconstruction network the $K$th iterate of proximal gradient descent $\bx^{(K)}$ starting from the initialization $\bx^{(0)} = 0$. Then we can turn this into a trainable reconstruction network by replacing all instances of the proximal operator $P(\cdot)$ with a trainable deep CNN $P_\theta(\cdot)$ mapping from images to images. In this approach the reconstruction network can be interpreted as learning a proximal operator. Any other free parameters, such as the step-size parameter $\eta$ can also be learned in training. 

The unrolling approach presented above was pioneered in \cite{Gregor2010LFA} in a sparse coding context. Unrolled versions of (proximal) gradient gradient descent for inverse problems in imaging are investigated in \cite{chen2017trainable,diamond2017unrolled,mardani2018neural,liu2019deep}. Other optimization algorithms that have been investigated in an unrolling context include the alternating directions method of multipliers (ADMM) \cite{sun2016deep}, primal-dual methods \cite{adler2018learned}, half-quadratic splitting \cite{schmidt2014shrinkage,zhang2017learning}, block coordinate descent \cite{ravishankar2017physics, ravishankar2018deep,chun2018deep}, alternating minimization \cite{aggarwal2018modl}, and approximate message passing \cite{borgerding2017amp,metzler2017learned}. Beyond unrolling of optimization algorithms, recent work \cite{gilton2019neumann} considers an unrolling strategy based on a Neumann series approximation to the solution map of Equation~\eqref{eqn:MAP}.
    \subsubsection{Train from $\by$'s only (measurements only)} \label{sec:1d}
    If both the forward model $A$ and the noise statistics are known, then the measurements themselves can be used as a proxy for the ground truth. In this case, it is possible to train reconstruction networks similar to those in Section~\ref{sec:1a} from the measurements alone, with an approporiate modification of the training loss function.
    This is known as self-supervision, and has been studied in~\cite{tamir2019unsupervised, yaman2019self} to learn autoencoders for estimating images $\bxstar$ from noisy measurements $\by$. Below we highlight a self-supervised approach based on Stein's Unbiased Risk Estimator (SURE).
    
    \paragraph{GSURE} In classical statistics, SURE~\cite{stein1981estimation} is a technique to compute the mean square error of a mean estimator, without access to the ground truth. 
    In order to understand how it can be used in deep learning for inverse problems, consider the denoising problem, where $\by = \bxstar + \bm{\varepsilon}$.
    Given a parametric class of estimators $\{ f_\theta \}_{ \theta \in \Theta}$ parameterized by $\theta\in \Theta$, SURE estimates the mean square error of $f_\theta$ given $\by$ as
    \[
    \E_{\bm\varepsilon} \left[\frac{1}{n}\| \bxstar - f_\theta(\by)\|^2\right] = \E_\eps\left[\frac{1}{n}\| \by - f_\theta(\by)\|^2\right] + 2\frac{\sigma^2}{n}\text{div}_{\by}\left(f_\theta( \by )\right) - \sigma^2 ,
    \]
    where $\sigma^2$ is the variance of $\bm\varepsilon$ and $\divg_{\by}(f_\theta(\by)) := \sum_{i=1}^n \tfrac{\partial f_\theta(\by)}{\partial y_i}$.
    Notice that computing the right-hand side of this equation does not require knowledge of $\bxstar$.
    
    If the estimators are differentiable with respect to $\theta$, then we can use gradient descent to learn a good estimator (i.e., $\theta^*$ are the parameters obtained by gradient descent, then the estimate of $\bxstar$ is given by $f_{\theta^*} (\by)$).
    This permits denoisers that are learned using noisy measurements alone. 
    SURE can be generalized to other forward models $A$ via GSURE~\cite{GSURE,metzler2018unsupervised}, by minimizing the following functional with respect to $\theta$:
    \[
    \E_{\vareps} \left[\frac{1}{n}\|P_A(\bxstar - f_\theta(\by))\|^2\right] = \E_{\vareps}\left[\frac{1}{n}\| P_A \bxstar \|^2 + \frac{1}{n}\| P_A f_\theta(\by)\|^2 - \frac{2}{n}f_\theta(\by)^T A^\dagger \by + \frac{2\sigma^2}{n}\divg_{\by}(f_\theta(\by)) \right],
     \]
    where $A^\dagger$ is the pseudoinverse of $A$ and $P_A = A^\dagger A$ is projection onto the row space of $A$.
    
    Notice that there is considerable freedom in choosing the function $f_\theta$. In particular, one can employ any of the reconstruction networks described in the supervised setting above.
    In~\cite{soltanayev2018training,metzler2018unsupervised,zhussip2019training}, the authors apply SURE to train the DnCNN~\cite{zhang2017beyond} and Learned Denoising-Based Approximate Message Passing (LDAMP) networks~\cite{metzler2017learned} for denoising and compressive sensing tasks.

\subsection{Forward Model Known Only at Test Time}\label{sec:Atest}

We now consider the case where the forward model $\cA$ is known only at test time, and one has access to representative samples of the ground truth during training.
The algorithms surveyed here have the property that after training a deep model once, the same deep model can be used for any forward model.
This is advantageous in situations where ground truth data is abundant, but training deep models for different forward models is expensive.

\subsubsection{Train from $\bx$'s only (ground truth only)} 
\label{sec:2c}
When presented with only ground truth data at train time, there are two popular approaches in the literature. The first learns a proximal operator, or denoiser, that can be used in an iterative reconstruction algorithm, while the second utilizes the ground truth training images to learn a generative prior.

    \paragraph{Learning a proximal operator or denoiser from data.} 
    The plug-and-play (PnP) ~\cite{venkatakrishnan2013plug} approach is a powerful method for  solving inverse problems using exising image denoising algorithms.
    The high level idea behind these methods is to use denoisers, such as BM3D~\cite{dabov2007image}, in place of proximal operators in iterative optimization algorithms such as ADMM~\cite{boyd2011distributed, parikh2014proximal}.
    The denoiser acts as a regularizer for the reconstruction, and ensures good reconstruction quality at each iteration of the algorithm, such that the final reconstruction matches the measurements and satisfies the prior defined by the denoiser. A closely-related approach
    \cite{romano2017little} called Regularization by Denoising (RED), proposed a general framework for PnP methods that can use deep neural networks as denoisers, by changing the functional used for regularization.
    Improved methods for training deep neural networks for PnP can be found in~\cite{meinhardt2017learning, zhang2017learning, gupta2018cnn}.
     
    Inspired by the success of Approximate Message Passing algorithms (AMP)~\cite{donoho2009message} for compressed sensing, Learned Denoising-Based Approximate Messaging Passing (LDAMP)~\cite{metzler2017learned} learns a denoiser which can be used in a variant of AMP. 
    Emprical results show that~\cite{metzler2017learned} can achieve state of the art reconstructions with a $100$-fold speed improvement over other state of the art methods.
    Additionally, LDAMP has a state evolution heuristic which can predict the mean square error of the reconstruction at each iteration.

    A similar approach is considered in~\cite{rick2017one}, where a denoiser is learned from data via adversarial training. 
    This denoiser is used as a proximal operator in the Alternating Direction Method of Multipliers (ADMM) algorithm~\cite{boyd2011distributed} to estimate $\bxstar$.     
    
    Notice that all of these approaches are flexible and can be used to solve general inverse problems, since training the denoiser is independent of any fixed forward model.

    \paragraph{Learning a generative prior from data.}
     A complementary approach to learning a proximal operator is to learn a model which is capable of generating new images based on the training samples.
     Compressed Sensing using Generative Models (CSGM)~\cite{bora2017compressed} demonstrated how deep generative models can be used for solving inverse problems.
     The first step of CSGM~\cite{bora2017compressed} is to train a generative model $G:\R^k \rightarrow \R^n, k\ll n,$ to capture the distribution of $\bx$, given training data. 
     This involves training a deep generative model, which can be trained using a varity of methods, such as adversarial training for GANs~\cite{goodfellow2014generative} or variational inference for VAEs~\cite{kingma2013auto}.
     Once a deep generative model $G$ is trained, the estimate of a measured image $\bxstar$ is obtained by solving the following optimization problem:
     \begin{equation}\label{eq:csgm}
     \hat{\bm z}:=\argmin_{\bm z \in \R^k} \|A G(\bm z) - \by\|^2,
     \end{equation}
     and the reconstruction is given by $G(\hat{\bm z})$.
     In words, we search in the latent space of the generative model $\R^k$ for a generated image that best explains the measurements. The optimization problem \eqref{eq:csgm} is non-convex and actually NP-hard~\cite{lei2019inverting}. CSGM~\cite{bora2017compressed} proposed solving this problem by starting from a random initalization $z_0 \in \R^k$ and performing 
     gradient descent (or ADAM~\cite{kingma2014adam}) to find the generated image that best fits the measurements. 
     
Similar ideas like projections on smooth manifolds and additional structure beyond sparsity in inverse problems have been studied in earlier signal processing work, e.g.~\cite{hegde2008random,baraniuk2009random,baraniuk2010model,eldar2009robust}. 
    Empirical results in~\cite{bora2017compressed} show that CSGM can
    achieve similar reconstruction quality using $5-10$ fold fewer measurements compared to sparsity-based LASSO methods. 
    
    CSGM~\cite{bora2017compressed} also generalized the theoretical framework of compressive sensing and restricted eigenvalue conditions~\cite{tibshirani1996regression,donoho2006compressed,bickel2009simultaneous,candes2008restricted} for signals lying on the range of a deep generative model. 
    For random subgaussian measurement matrices $A$, 
    a condition called the Set Restricted Eigenvalue condition (SREC) \cite{bora2017compressed}, can be used to show the following two results:
    \begin{itemize}
    	\item if $G$ is an $L-$Lipschitz function, $m=O(k\log \frac{Lr}{\delta})$ measurements suffice to guarantee\\ ${\|G(\hat{\bm z}) - \bxstar\|\leq 6 \min_{\bm z:\|\bm z\|\leq r} \|G(\bm z) - \bxstar\| + \delta}$.
    	\item if $G$ is a $d-$layered feedforward neural network with piecewise linear activation functions, then $m=O(kd\log n)$ measurements suffice to guarantee $\|G(\hat{\bm z}) - \bxstar\|\leq 6 \min_{\bm z\in\R^k} \|G(\bm z) - \bxstar\| $.
    \end{itemize} 
    More recently, lower bounds~\cite{kamath2019lower,liu2019sample} 
    established that these numbers of measurements are order optimal.
Further, \cite{kamath2019lower} demonstrated that deep generative models can produce all $k$-sparse signals, hence modeling structure with DGMs is a strict generalization of sparsity.
    The subgaussian assumptions on $A$ were relaxed in~\cite{jalal2020robust}, which further proposed a new algorithm which is robust to heavy tailed noise and arbitrary outliers.
    Further, asymptotically optimal results can be found in~\cite{jalali2019solving, aubin2019exact}.

    These results guarantee that the optimium $\hat{\bm z}$ of \eqref{eq:csgm}
    will be close to the best possible representation that the generative model can achieve. Unfortunately, actually finding this optimum is computationally hard~\cite{lei2019inverting} so it not known what can be achieved provably in polynomial time, despite the excellent empirical performance of gradient descent\footnote{Empirically, gradient descent inversion works well for medium-sized generative models like DCGAN but has not been very effective in inverting bigger generators like BigGAN~\cite{brock2018large}, see \cite{daras2019your} for a discussion.} 
    Hand et al.~\cite{hand2017global} made important theoretical progress assuming that the weights of the generative model $G$ are random and independent. For random weights, and further assuming that each layer of the generative model grows by a logarithmic factor,~\cite{hand2017global} proved that the objective \eqref{eq:csgm} has only two local minima and can be optimized by gradient descent. An analysis of projected gradient descent for this problem was given by~\cite{shah2018solving}, while ADMM methods were proposed and analyzed in~\cite{latorre2019fast}. Analyzing gradient descent inversion for generative models that do not expand logarithmically per layer  
    (as postulated by~\cite{hand2017global}) remains as an open problem. The least squares objective in \eqref{eq:csgm} can be seen as a projection on the range of a generator, and was also independently proposed in~\cite{mardani2017deep, tripathi2018correction, anirudh2019mimicgan}. 
    
    The CSGM approach has been generalized to tackle different inverse problems, algorithms for decoding, and to other assumptions on the generative model.
    Examples of different inverse problems include phase retrieval~\cite{hand2018phase, aubin2019exact   }, blind deconvolution~\cite{asim2018blind}, 
    geophysical seismic imaging~\cite{mosser2020stochastic},
    bilinear estimation~\cite{asim2018solving}, and 1-bit compressed sensing~\cite{qiu2019robust}.
    \cite{dhar2018modeling, kabkab2018task} propose improvements to the objective function in \eqref{eq:csgm}.
    Alternate algorithms for decoding including ML-VAMP~\cite{fletcher2018inference, fletcher2018plug} and Surfing~\cite{song2019surfing}.
    The results in~\cite{lindgren2020conditional} provide uncertainty quantification for the reconstruction.

    While trained generative models are a powerful tool for solving inverse problems, training them can be challenging since they require massive datasets and a long training time.
    Surprising results show that \emph{untrained generative networks} can solve this difficulty, and we review this line of work in~\Cref{sec:open}.

\subsubsection{Train from $\by$'s only (measurements only)}\label{sec:2abd}
 Having $\by$'s among the training data suggests that there is a fixed $\cA$ generating training samples. While work in this setting should provide few technical challenges, we are unaware of application domains in which this paired data would be available yet $\cA$ would be wholly unknown except at test time.

\subsection{Forward Model Partially Known}\label{sec:Apartial}
    
    We now consider inverse problems where the forward operator is partially known.
    This can occur, for example, when the forward model is parametric and we know either the distribution of or sufficient statistics about the parameters.
    
   \subsubsection{Train from  $(\bx,\by)$ pairs (paired ground truth and measurements)}
   \label{sec:3a}
    
    In general, knowledge of $\cA$ arises from a mathematical model of an imaging system or careful calibration procedures. In either case, we typically only know an approximation of $\cA$. In general, these inaccuracies can complicate the reconstruction process, but when we have real-world training observations of the form $(\bx,\by)$, then we can expect those samples to reflect the true $\cA$. As a result, training a deep neural network to perform reconstruction can leverage the partial knowledge of $\cA$ to perform some approximate inversion of the measurement process while using the training data to learn to remove ``artifacts'' and compensate for inaccuracies in the model. See the illustration in Figure~\ref{fig:residual}.

   \subsubsection{Train from unpaired $\bx$'s and $\by$'s  (unpaired ground truth and measurements)}
   \label{sec:3b}
    In certain cases, one has access to unpaired samples of the ground truth and measurements.
    That is, if $\bxstar_i, \by_i$ denote the $i^{th}$ training sample of the ground truth and measurement, then $\bxstar_i, \by_i$ follow the marginal distributions of $\bxstar, \by$, without following the joint distribution of $(\bxstar, \by)$.
    This can occur for example, if one has clean MRI scans as ground truth, and MRI scans with motion blur as measurements, without any pairing between the clean and blurry scans. 
    
    Models like CycleGAN~\cite{zhu2017unpaired} are well-suited for this situation, as they can learn forward and backward mappings between the image and measurement domain, given unpaired samples of images and measurements.
    This idea has been explored in~\cite{armanious2019unsupervised}, for removing motion blur from MRI scans, as well converting PET scans to CT scans.
    A similar idea was explored for MRI by~\cite{quan2018compressed}, although in this case the forward operator is assumed to be a subsampled Fourier transform.
    
    We briefly describe the original CycleGAN algorithm, as its extension to inverse problems can be derived with domain specific modifications.
    Let $p_{\bx}, p_{\by},$ denote the distributions over $\bx, \by$.
    CycleGAN aims to learn two generative models $F: \mathcal{X} \rightarrow \mathcal{Y}, G: \mathcal{Y} \rightarrow \mathcal{X}$, where $\mathcal{X, Y}$ are respectively the image and measurement domain. 
    Since $G, F,$ need to be trained with unpaired $\bx$'s and $\by$'s, one way to create a joint distribution between $\bx, \by$ is to make $G,F$ approximate inverses of each other.
    That is, for all $\by \in \mathcal{Y},  F(G(\by)) \approx \by,$ and for all $\bx \in \mathcal{X}, G(F(\bx)) \approx \bx.$
    In~\cite{zhu2017unpaired}, this requirement is satisfied by introducing the \emph{cycle consistency loss}, defined as
    \[
    \mathcal{L}_{cyc} (G,F)= \E_{p_{\bx}} [\| \bx - G(F(\bx))\|_1 ] + \E_{p_{\by}} [\| \by - F(G(\by)) \|_1].
    \]
    By adding this cycle loss to individual adversarial losses for $G$ and $F$, they can be simultaneously trained.
    Once they are trained, $F,G$ can be used to map from the image domain to the measurement domain, and vice versa.
    For example, if $\by$ is an MRI scan with motion blur, then $G(\by)$ will remove the blur present in $\by$.

	\subsubsection{Train from $\bx$'s only (ground truth only)}\label{sec:3c}
	Generative priors~\cite{kingma2013auto, goodfellow2014generative} have been successfully applied in many inverse problems~\cite{bora2017compressed}, and are a good option when one has access to samples of $\bx$.
	These priors have also found success in problems like blind deconvolution~\cite{asim2018blind}, where $\cA (\bx) = \bx \circledast A, $ and the distribution of $A$ is known, but we do not know the exact blurring kernel that produced the measurements. 
	In~\cite{asim2018blind}, it is assumed that there exists two generative models given by $G_{A}, G_{\bx}$: the output distribution of $G_{A}$ captures the distribution of $A$, whereas $G_{\bx}$ captures the distribution of $\bx$.
	
	Given $G_{A}, G_{\bx}$, and measurements $ \by = \bxstar \circledast A ^* + \bm \varepsilon$, one can recover the ground truth $\bxstar$ and the blurring kernel $A^*$ by solving the following optimization problem:
	\begin{align}
	    \hat{\bm z}_{\bx}, \hat{\bm z}_{A} = \argmin_{\bm z_{\bx} \in \R^k, \bm z_{A} \in \R^k} \| y - G_{\bx}(\bm z_{\bx}) \circledast G_{A}(\bm z_{A}) \|^2.
	\end{align}
	Once this is solved, the estimates for $\bxstar, A^*,$ are given by $G_{\bx}(\hat{\bm z}_{\bx}), G_{A}(\hat{\bm z}_{A})$.
	
	This approach was generalized to blind demodulation by~\cite{hand2019global}.
	\cite{hand2019global} also provide theoretical guarantees on the loss landscape of the above objective, and show that it can be minimized by gradient descent.
	
	Another approach is DeblurGAN~\cite{kupyn2018deblurgan}, in which a GAN is trained end-to-end using blurry images as input.
	During training, clean images are synthetically blurred, and the GAN must learn to generate crisp images given blurry images.
	This produces cripser images, but it can be expensive, since a minor change in the distribution of $A$ would require retraining the GAN.

	\subsubsection{Train from $\by$'s only (measurements only)} 
	\label{sec:3d}
	
	Learning from measurements alone is a hard task, which is further compounded by the difficulty of only having partial knowledge of the forward model.
	In order to tackle this problem, it is commonly assumed that the forward operator has an underlying distribution, and we have knowledge of its statistics.
	There are two popular ways of solving this problem: one is a ``supervised'' approach, and another is via adversarial training. 
	We first explore the supervised approach.
	
	\paragraph{Noise2Noise.}
	Noise2Noise~\cite{lehtinen2018noise2noise} learns a neural network $f_\theta : \R^m \rightarrow \R^n$ that accepts noisy measurements as input and produces clean samples as output.
	The training of $f$ is reminiscent of supervised training, except it does not actually need ground truth.
	In order to train $f$,~\cite{lehtinen2018noise2noise} assume
	\begin{itemize}
	    \item The training data consists of $(\tilde{\bm x},\by)$ pairs, where $\tilde{\bm x}$ is a noisy version of $\bxstar$, and $\by = \cA(\bxstar) + \bm\varepsilon$.
	    \item The samples $\tilde{\bx}$ satsify $\E [ \tilde{\bm x} | \by] = \bxstar$.
	\end{itemize}
    Given this dataset, the learned neural network is $f_{\theta^*},$ where
    $ \theta^* = \argmin_{\theta} \E [ \|f_\theta( \by ) -\tilde{\bm x} \|^2]. $
	The theoretical argument for Noise2Noise is based on the assumption $\E [\tilde{\bx} | \by] = \bxstar$.
	This allows $f$ to be trained from contaminated samples $\tilde{\bx}$, without access to the clean ground truth.
	In theory this should require multiple $\tilde{\bx}$ for each $\bxstar$, but~\cite{lehtinen2018noise2noise} observe that one sample suffices. 
	An important benefit of this approach is that it does not need explicit knowledge of the parameters or distribution of $\cA$.
	While Noise2Noise does not need the ground truth, it still requires $\tilde{\bx}$, which is a noisy proxy of $\bxstar$.
    We now explore an alternative approach which weakens this assumption.	
	
	\paragraph{Adversarial training.} 
	Adversarial training has emerged as a powerful technique for learning high dimensional distributions that are hard to describe.
	When $\cA$ follows a parametric distribution, AmbientGAN~\cite{bora2018ambientgan} demonstrates how adversarial training can learn from measurements alone.
    With a slight abuse of notation, let $\by, \bxstar, \cA,$ respectively denote random variables associated with the measurements, ground truth, and forward model. 
    Similarly, let $P_{\by}, P_{\bxstar}, P_{\cA}$ denote their probability distributions.
    Given samples from $P_{\by}$, and assuming that it is easy to sample parameters of the forward model, AmbientGAN learns the distribution $P_{\by}$ by optimizing the following objective:
	\[
    \min_{G}\max_{D} \E_{\by}\left[\log(D(\by))\right] - \E_{\bz,\cA}\left[\log(1-D(\cA (G(\bz))))\right],
    \]
    where $G:\R^k\rightarrow\R^n, D:\R^n \rightarrow [0,1]$ with $k \ll n$, and $\bz \in\R^k$ is a random latent variable which can be easily sampled, for example i.i.d.~Gaussian or i.i.d.~uniform. 
    The intuition for this approach is similar to that of traditional GANs~\cite{goodfellow2014generative}.
    In traditional GANs, the discriminator $D$ must learn to distinguish between the distribution $P_{\bxstar}$ and $P_{G(\bz)}$, whereas in AmbientGAN, the discriminator must learn to distinguish between $P_{\by}$ and $P_{\cA G(\bz)}$.
    Under certain regularity conditions on the distributions $P_{\cA}, P_{\bxstar}$,~\cite{bora2018ambientgan} show that the distribution $P_{\bxstar}$ can be exactly recovered.
    
     Once the AmbientGAN is trained, it can be used for inference: for a new $\cA,\by$, the reconstruction $\hat{\bx} = G(\hat{\bz})$ can be obtained by solving the constrained least squares problem \eqref{eq:csgm}.
    Note that if the ground truth distribution has been exactly learned by AmbientGAN, $\cA$ need not follow any distributional assumptions in the inference phase.
    
	While AmbientGAN has nice theoretical properties, it can be computationally expensive, since it requires running an optimization procedure  at test time.
	A more direct solution is to train a network $\tilde{G}$ which accepts the measurements as input and outputs a possible reconstruction. 
	\cite{pajot2018unsupervised} is one such approach, where the reconstruction $\tilde{G}(\by)$ is ideally the MAP estimate of $\bxstar$.
	Similar ideas have been explored in~\cite{sonderby2016amortised}, although in~\cite{sonderby2016amortised} there is no stochasticity in $\cA$.

\subsection{Unknown Forward Model}\label{sec:Aunknown}
In some cases the forward model may be entirely unknown, misspecified, or computationally intractible to use in training and testing. If this is the case, then one is essentially limited to the supervised setting, \ie learning must take place with matched image and measurement pairs.

\subsubsection{Train from  $(\bx,\by)$ pairs (paired ground truth and measurements)}
\label{sec:4a}
\begin{figure}
    \centering
    \includegraphics[width=.5\columnwidth]{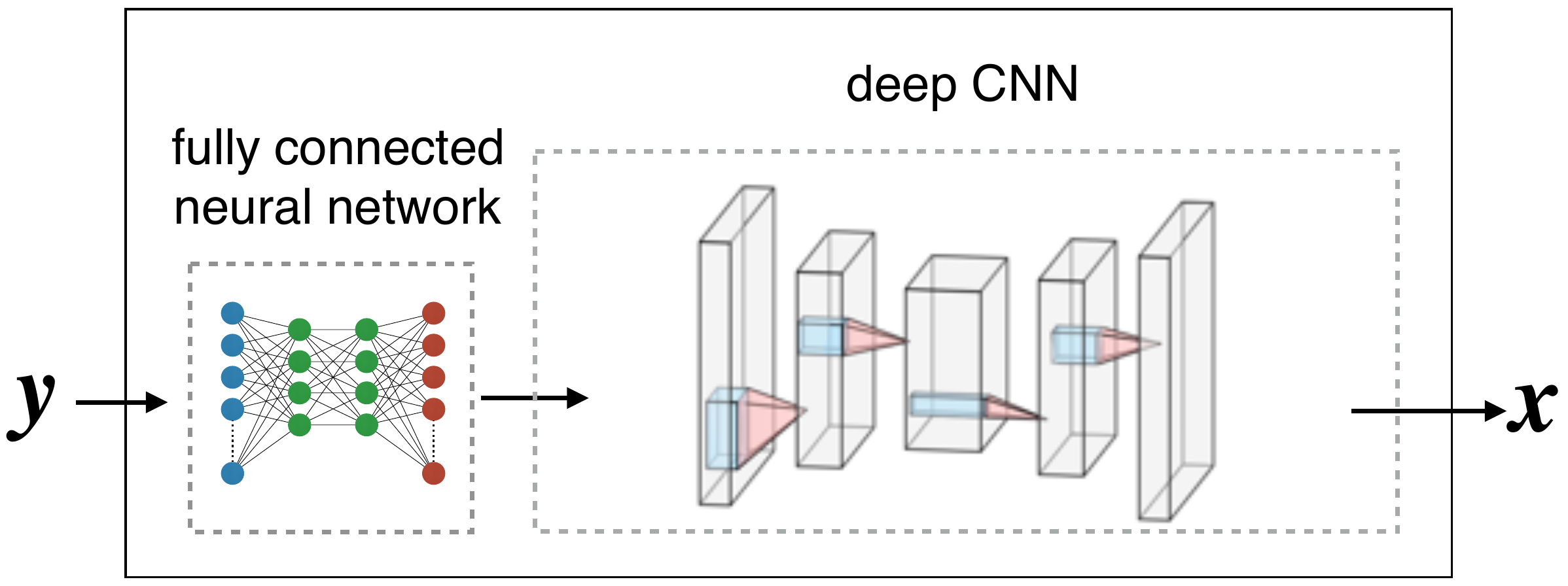}
    \caption{When the forward model $A$ is unknown but sufficiently many training samples are available, learning a reconstruction map is still possible using deep neural networks, as proposed in the AUTOMAP framework \cite{zhu2018image}.}
    \label{fig:agnostic}
\end{figure}

Assuming that one only has access to image and measurement pairs $(\bx,\by)$ without knowledge of the forward model the simplest approch is to treat reconstruction map $\by \mapsto \bx$ as a ``black box'' that can be well-approximated by conventional neural network architectures with the appropriate input and output dimensions, as illustrated in Figure~\ref{fig:agnostic}. This is the approach taken in \cite{zhu2018image}, which proposed the automated transform by manifold approximation (AUTOMAP) framework. In this framework, the reconstruction network $f_\theta$ is modelled as a map between a low-dimensional ``measurement manifold'' $\mathcal{Y}$ and an ``image manifold'' $\mathcal{X}$ embedded in high-dimensional Euclidean space: $f_\theta = \phi_{\bx}\circ g \circ \phi_{\by}^{-1}$, 
where $\phi_{\by}^{-1}$ maps Euclidean space to intrinsic coordinates in $\mathcal{Y}$, $g$ is a diffeomorphism between $\mathcal{Y}$ and $\mathcal{X}$, and $\phi_{\bx}$ maps from instrinic coordinates in $\mathcal{X}$ to Euclidean space. To approximate this idealized mapping, $\phi_{\by}^{-1}$ is then parameterized as a sequence of fully connected neural network layers, while $g$ and $\phi_{\bx}$ are parameterized as a sequence of CNN layers. Note that while the experiments in \cite{zhu2018image} used knowledge of the forward model $A$ to generate training data from fully sampled images, in principle the approach should succeed without access to $A$.

\subsubsection{Unsupervised approaches}\label{sec:4bcd}
If $\cA$ is entirely unknown, then there are limited options without paired $(\bx,\by)$ training samples, as these pairs are our only mechanism for understanding anything about $\cA$. In order to make make this problem identifiable, it is necessary to have some additional information about $\cA$.

\section{Key Tradeoffs}

\subsection{Sample Complexity vs.\ Generality}\label{sec:complexity}

In many of the unsupervised learning approaches we discussed above, training takes place independently of the forward model $A$. This includes compressed sensing with generative models and iterative plug-and-play reconstruction with a denoising autoencoder. In these cases, one learns a generative model or denoising autoencoder using only a collection of training images, which does not require knowledge of $A$. The advantage of this approach is that once training has taken place, the learned generative model or denoising autoencoder can be used for {\em any} forward model, so we do not need to re-train a system for each new inverse problem. In other words, the learning is {\em decoupled} from solving the inverse problem, resulting in high generality.

However, the generality of the decoupled approach comes with a high price in terms of sample complexity. To see why, note that learning a generative model or a denoising autoencoder fundamentally amounts to estimating a full \emph{prior} distribution over the space of images; let us denote this distribution as $p(\bx)$. 
\begin{figure}[ht]
    \centering
    \subfloat[]{\includegraphics[width=.2\linewidth]{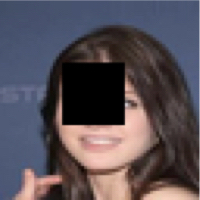}} ~
    \subfloat[]{\includegraphics[width=.2\linewidth]{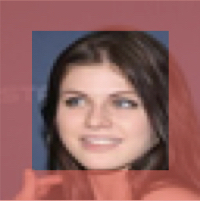}}
    \caption{Sample complexity in learning to solve inverse problems. (a) An example of an inpainting problem in which the goal is to estimate the missing (black) pixels in the center of the image. If we do not know ahead of time which pixels may be missing, then we must learn (perhaps implicitly) the distribution over all possible images, $p(\bx)$. (b) If we know at training time which pixels are going to be missing, this information can be used to reduce sample complexity. For instance, if we know the missing pixels will be located as they are in (a), then the red highlighted pixels in (b) are irrelevant to the inpainting task and we do not need to learn the distribution of pixels values in those regions. Rather, we must only learn the {\em conditional} distirbution $p(\bx| \by)$, which can require significantly fewer training samples.}
    \label{fig:inpainting}
\end{figure}
Thoroughly understanding the space of images of interest is important if our learned regularizer is to be used for linear inverse problems of which we are unaware during training. However, when we know at training time what $A$ is, then we only need to learn the {\em conditional} distribution $p(\bx | \by)$ where $\by = A\bx + \bm{\eps}$ . For example, consider an inpainting scenario in which we only observe a subset of pixels in the image $\bx$. Rather than learn the distribution over the space of all possible images, we only need to learn the distribution over the  space of missing pixels conditioned on the observed pixels, as in Figure~\ref{fig:inpainting}. Of course, if we know the forward model $A$ and the statistics of the noise $\bm \eps$, then $p(\bx | \by)$ can be calculated from $p(\bx)$ and $A$ using Bayes' law. However, such an approach is not always the most sample-efficient. 

For instance, imagine our images have $d$ pixels and the distribution $p(\bx)$ lies in a Besov space with a smoothness parameter $\alpha$, where larger $\alpha$ implies smoother functions that are easier to estimate \cite{devore1993constructive}. Then the $L^2$ density estimation error scales like $N^{-\frac{\alpha}{2\alpha+d}}$, where $N$ is the number of training samples \cite{donoho1996density,delyon1996minimax,lafferty2008minimax}. In contrast, conditional density estimation errors scale like $N^{-\frac{\alpha'}{2\alpha'+d'}}$, where $\alpha'$ is the smoothness of the conditional density and $d'$ is the number of pixels on which the conditional density depends \cite{efromovich2007conditional,bertin2016adaptive} (\ie the number of pixels not covered by the red overlay in Figure~\ref{fig:inpainting}) . In many scenarios $\alpha' > \alpha$ and $d' \ll d$, meaning that conditional density estimation can achieve much smaller errors with many fewer training samples than strategy of first estimating the full density and then calculating the conditional density based on this higher-error estimate.

The key point is that decoupled approaches (implicitly) require learning a full prior $p(\bx)$ whereas a method that incorporates $A$ into the learning process has the potential to simply learn the conditional density $p(\bx | \by)$, which often can be performed accurately with relatively less training data.

\subsection{Reconstruction Speed vs.\ Accuracy}

In many inverse problems in imaging, the computational bottleneck in traditional reconstruction algortihms comes from applying the forward model $A$ (or its adjoint $A^\T$). This is a perpetual challenge in applications such as medical image reconstruction. As a result, approaches based on iterative optimization with a large number of step sizes can be quite time-consuming 
\cite{rudin1992nonlinear,venkatakrishnan2013plug,wang2008new,romano2017little}
even though the resulting reconstruction may be highly accurate.
Deep learning provides an opportunity to reduce the computational burden of model-based iterative reconstruction methods. 

For instance, consider iterative reconstruction methods. Each iteration typically requires at least one application of the forward model operator $\cA$ and its adjoint, and these calculations can be the primary computational burden of the method. Reducing the number of iterations can therefore dramatically reduce reconstrution time. In classical methods, like iterative total-variation regularized reconstruction, we have few mechanisms for controlling the number of iterations since the methods need to run to convergence to yield accurate results. However, empirical results have shown that deep-learned based approaches can achieve comparable accuracies with far less computation. For instance, consider the unrolled optimization methods described above; there the number of blocks is akin to the number of iterations, and by fixing the number of blocks and then learning a regularizer within this framework, we essentially learn a reconstruction method that is adapted to a small number of iterations. Specialized unrolling approaches incorporating preconditioners can reduce the number of blocks and further increase the speed.

\section{Caveats/Beware/Failure Modes}
\label{sec:caveats}

So far we have reviewed some of the exciting breakthroughs that have been made possible through deep learning. In this section we view these algorithms through a more critical lens, in order to understand their current limitations and failures. This raises important research questions that we must address before we can hope to apply deep learning in real world applications of inverse problems.

\begin{figure}[ht!]
    \centering
    \subfloat[]{\includegraphics[width=.5\linewidth]{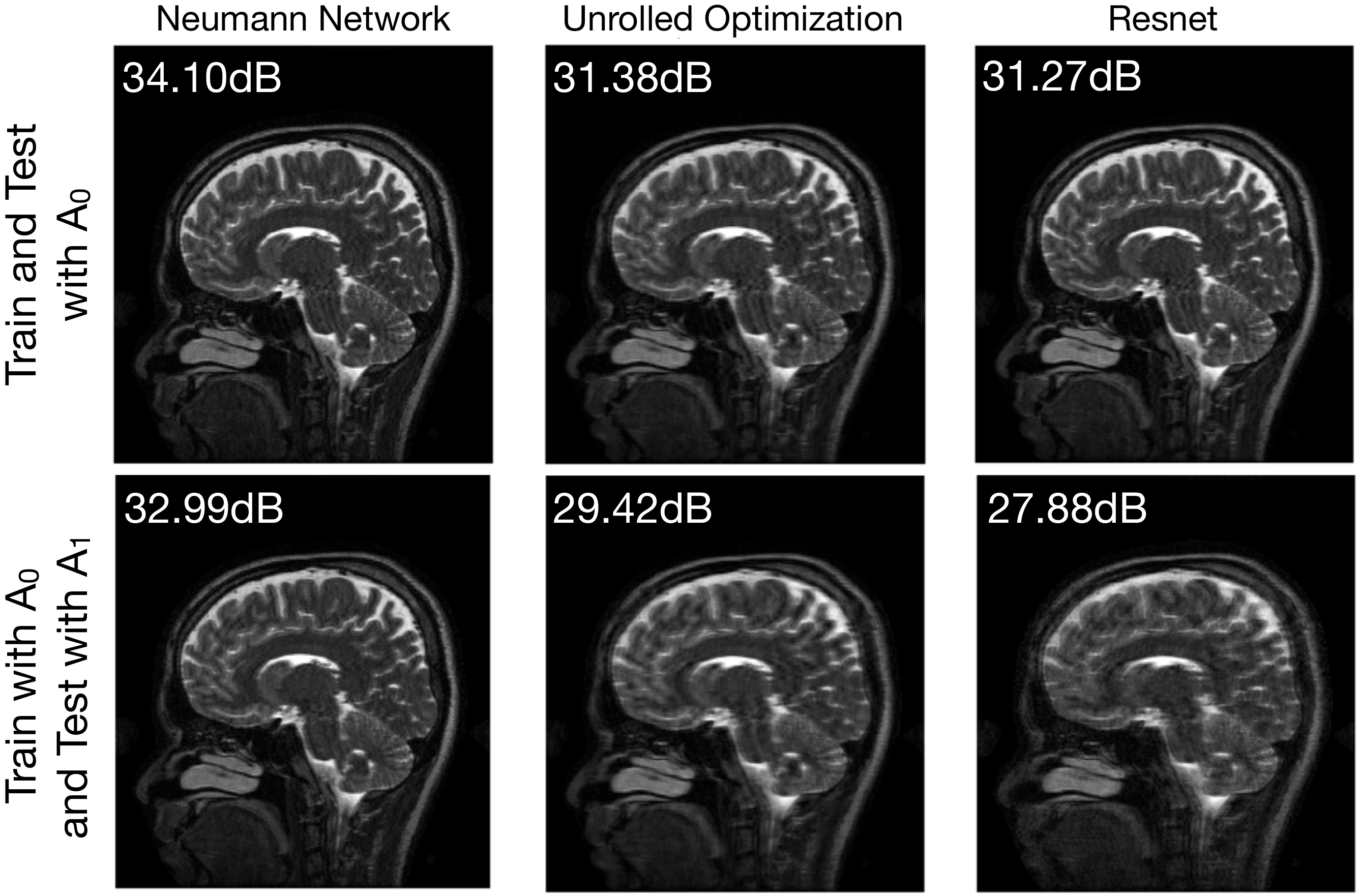}}~
    \subfloat[]{\includegraphics[width=.5\linewidth]{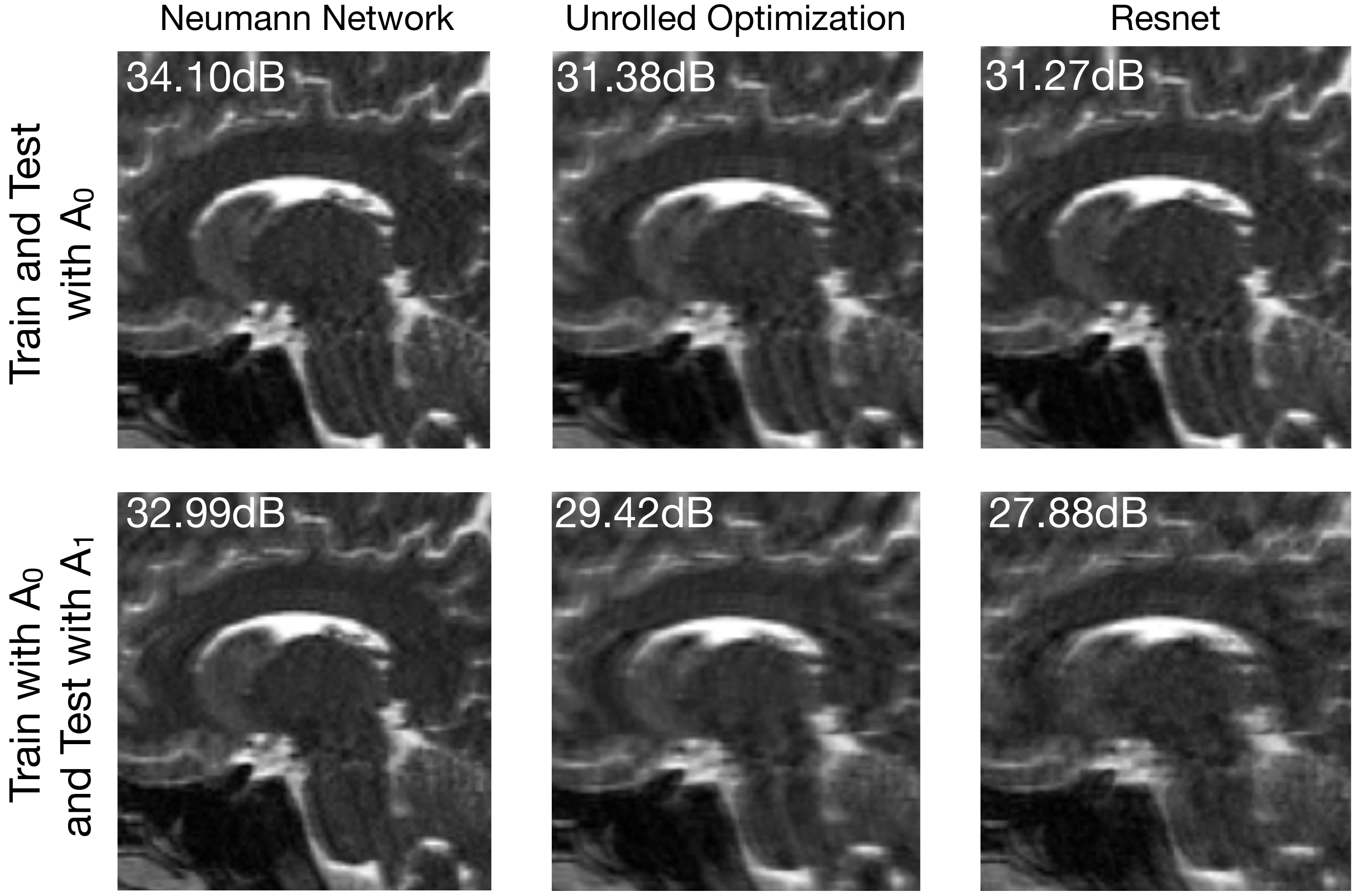}}
    \caption{Robustness to forward model perturbations. (a) Reconstructions produced by Neumann networks \cite{gilton2019neumann}, unrolled optimization \cite{diamond2017unrolled}, and a residual autoencoder \cite{mao2016image}. The first row shows reconstructions when the true forward model is used during training and the same forward model is true and used during testing. The second row shows reconstructions when the true forward model is used during training and is also used at test time even though a different forward model generated the test data.  (b) Same as (a), but zoomed into a smaller region to help show details. This example illustrates the shortcomings of learning to reconstruct for a specific forward model $A$ and also how different learning frameworks can be more or less sensitive to perturbations of $A$. The tradeoffs between network architecture and robustness are not well understood and an active area of research.}
    \label{fig:robust_forward}
\end{figure}

\paragraph{Robustness to different forward model at test time than at train.}  In some settings, the forward model used during training is different from the forward model used during testing. For example, imagine learning to reconstruct MRI images for a scanner at one clinic and then attempting to use that learned algorithm to reconstruct MRI images for a (subtly different) scanner at another clinic. The different methods described in Section~\ref{sec:taxonomy} will have different degrees of robustness to perturbations in the forward models between training and testing. This is illustrated in Figure~\ref{fig:robust_forward} for a few representative methods. 

A related model mismatch issue arises in the discretization of forward models and learned reconstruction
algorithms. For example, in order to generate training data many supervised learning methods commit the ``inverse crime'' \cite{wirgin2004inverse} by assuming the forward model and ground truth images are discrete, when in fact they are defined in a continuous domain. This can lead to undesirable artifacts at test time, such as Gibb's ringing artifacts in MRI \cite{guerquin2011realistic}. Learning-based approaches by themselves cannot resolve this issue, and need to be properly modified in order to recover artifact-free images at test time by, e.g., incorporating post-processing steps.

\begin{figure}[ht!]
    \centering
    \includegraphics[width=.6\linewidth]{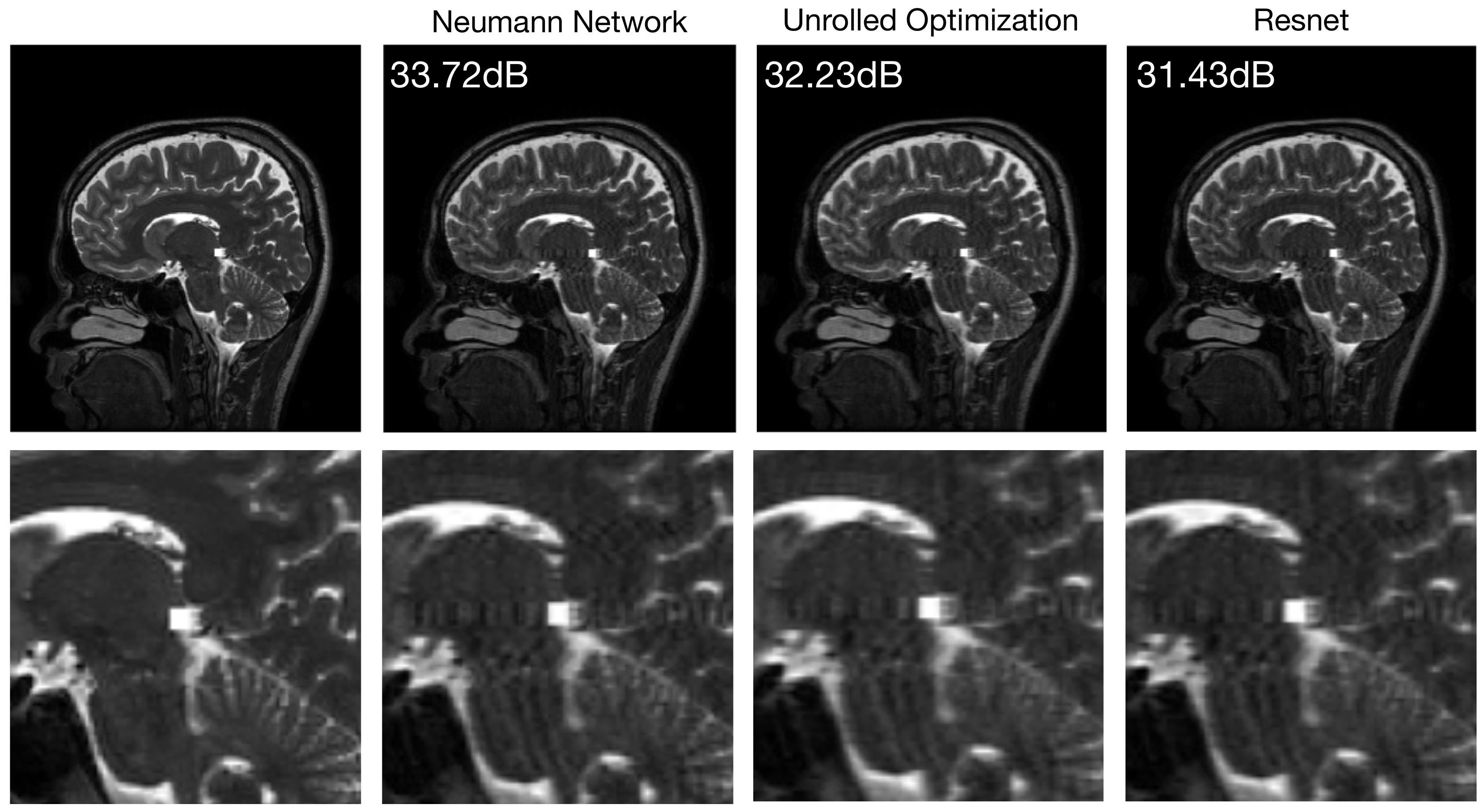}
    \caption{Robustness to image features not represented in training set. Three different reconstruction methods -- Neumann networks \cite{gilton2019neumann}, unrolled optimization \cite{diamond2017unrolled}, and a residual autoencoder \cite{mao2016image} -- were trained on an MRI training set from \cite{aggarwal2018modl}. They are then applied to a test image that corresponds to an MRI image with a small square inserted near the center, as shown in the far left column. This type of square feature was not present in the training set. The three methods do recover the feature, but also produce non-trivial artifacts to its left and right. Both original (top) and zoomed (bottom) images are shown.}
    \label{fig:stability}
\end{figure}

\paragraph{Recovering features not represented by training data.} The central assumption underlying all machine learning based image reconstruction methods is that the training data is representative of what we might see at test time. In some applications, such as medical imaging, it is unclear to what extent that assumption holds. One might imagine patients with unusual geometries in their anatomy or tumors that are not reflected by the training set \cite{antun2019instabilities,gottschling2020troublesome}.  The ability of learned reconstruction methods to faithfully reconstruct such features remains poorly understood and can vary from method to method, as illustrated in Figure~\ref{fig:stability}.

\paragraph{Difficult to interpret.} 
A side effect of the flexibility and power offered by deep learning models is that they are difficult to analyze and interpret. 
Hence we currently have a very poor understanding of some methods that provide state of the art results. 
For example, deep image prior~\cite{ulyanov2018deep} and related methods~\cite{heckel2018deep, van2018compressed} provide very surprising results -- they do not require any training data, but are competitive with methods that leverage knowledge from large datasets. 
The current hypothesis for their success is that convolutional models are biased towards smooth signals, and~\cite{heckel2019denoising} provide preliminary theoretical arguments for this hypothesis. 
However, a solid theoretical framework for analyzing these models remains open. 
A similar argument can be made against supervised models, wherein we do not understand the training phase well enough to analyze the predictions made at inference time. 

\paragraph{Creation of artifacts.} There has been significant progress in generative modeling over the last few years, and the perceptual quality of generated images is almost lifelike~\cite{brock2018large}.
Early GANs struggled with images containing complicated semantic structure, but modern GANs have been able to overcome this issue.
Despite the recent progress in developing better generative models, the generated images contain many artifacts and distortions.
Deep learning models for medical imaging that directly map from measurements to images are also somewhat contentious. 
Deep learning has an incredible ability to generate realistic looking images, even when the features in the image are not actually present~\cite{gottschling2020troublesome}. 
These artifacts can be problematic if the reconstructions are used for downstream tasks such as classification of tumors.

\paragraph{Failure modes may be hard to recognize.}
    As an illustrative example, consider the case of compressed sensing using generative models~\cite{bora2017compressed}.
    The decoding algorithm always returns an image within the range of the generative model, which by design will have high perceptual quality.
    This is a positive when the data is well behaved and does not have outliers.
    Now consider a case where the forward operator is heavily underdetermined or data has outliers.
    In this case, the algorithm in~\cite{bora2017compressed} will return a high quality image even if it has failed~\cite{jalal2020robust}.
    In contrast, handcrafted algorithms Lasso would simply return a non-sparse signal.
    This failure mode is easy to recognize for Lasso, but if we are using a generative model, we may completely miss the failure.
    This prompts the need for algorithms that are either robust to outliers~\cite{jalal2020robust}, or algorithms that can provide uncertainty quantification for their reconstructions such as in~\cite{lindgren2020conditional}.

\section{Open Problems and Future Directions}
\label{sec:open}

\paragraph{Control over forward model design.}
In many applications one has some degree over the measurement process. For instance, one can select which locations in k-space are sampled in an MRI scan or which DMD patterns are applied in a single-pixel-camera. While this problem has long been tackled using heuristics like variable density sampling~\cite{lustig2007sparse}, deep learning provides a mechanism to optimize the sampling pattern in a more principled way. One need only make the measurement model a trainable parameter that can be optimized with training data~\cite{mousavi2017deepcodec}. This idea can been used to do things like design illumination patterns for microscopes~\cite{hershko2019multicolor,horstmeyer2017convolutional,kellman2019physics,kellman2019data}.

Taking this idea one step further, deep learning can even be used to {\em design} physical systems. A series of recent works have modeled cameras as differentiable optical systems and then used back-propagation to design specialized optical filters and diffractive optical elements for improved demosaicing~\cite{chakrabarti2016learning}, color imaging and demosaicing~\cite{chakrabarti2016learning}, super-resolution and extended depth-of-field imaging~\cite{sitzmann2018end}, monocular depth estimation~\cite{chang2019deep,wu2019phasecam3d}, high dynamic range imaging~\cite{metzler2019deep}, and single-lens wide-field-of-view imaging~\cite{peng2019learned}.

\paragraph{Extensions to other application domains.} This tutorial has focused on inverse problems in imaging, but inverse problems abound in many different settings, including estimating boundary conditions for partial differential equations, estimating molecular structure from multi-modal measurements, radar, geophysics, and more. Many of the central themes of this tutorial, including understanding what must be known about the forward operator or the nature of the training data needed for various algorithms to be viable translate to these other settings, while the specifics of how to choose the neural network architecture or tradeoffs among different algorithms remain open questions. Furthermore, the discussion in this paper focuses on settings in which everything that is known about the physical setting of the inverse problem may be encapsulated in the forward model $\cA$; in more general settings, in which we may have access to additional physical contraints or side information, there is an opportunity to develop frameworks for  incorporating this knowledge. 

\paragraph{Unlearned Methods.}
Deep Image Prior (DIP)~\cite{ulyanov2018deep} is an algorithm which uses \emph{untrained generative models} for image reconstruction.
Given measurements $\by$ and the forward operator $\cA$, DIP initializes a generative network $G_\theta:\R^k\rightarrow\R^n$ with a fixed random input vector $\bz\in \R^k$, and optimizes over the network weights $\theta$.
The reconstruction $\bxhat$ is given by $G_{\theta^*}(\bz)$, where 
\begin{equation}
   \theta^* = \underset{\theta}{\argmin}\|\cA(G_\theta(\bz)) - \by \|^2.\label{eq:dip}
\end{equation}
Experimental results in~\cite{ulyanov2018deep} show that DIP is competitive with state of the art algorithms.
This result is surprising, since DIP requires no training data, and only requires measurements from a single sample.
Additionally, the number of weights in the generative network often exceeds the number of pixels in the image.
This implies that DIP should be able to find a set of weights that can fit any image, including Gaussian noise, and this is observed empirically in~\cite{ulyanov2018deep}.
In order to avoid this,~\cite{ulyanov2018deep} uses early stopping as a regularizer.
For example, if gradient descent is used to solve~\cref{eq:dip}, then early stopping means that gradient descent must be stopped before it converges.
An intuitive explanation for the success of DIP is that convolutional neural networks are biased towards smooth, ``natural'' images, and hence smooth components of an image will be reconstructed before the noisier components in the measurements.
Results in~\cite{van2018compressed, jagatap2019algorithmic, jagatap2019phase} generalize the results in~\cite{ulyanov2018deep}.

Deep Decoder~\cite{heckel2018deep} is a related algorithm which fixes issues encountered in DIP.
The Deep Decoder is an underparameterized network which is competitive with DIP and \emph{does not } require early stopping. A preliminary result in~\cite{heckel2018deep} shows that a single layer Deep Decoder will not fit Gaussian noise, and further analysis can be found in~\cite{heckel2019denoising}.
Beyond these preliminary insights, untrained generative models remain  poorly understood and theoretically surprising. A proper theoretical framework for understanding when they work, how to properly regularize them and how to measure their complexity remain significant open problems. 

\paragraph{Transfer learning.} In many settings, such as medical imaging, we may have only limited quantities of training data; in other settings, such as astronomy, we may not have access to any ``real'' training images but can generate simulated training data. In these settings, we face the challenge of leveraging data from a different application domain or from simulations to improve inverse problem solvers in our target domain. This challenge is generally referred to as ``transfer learning'' or ``domain adaptation'' \cite{raina2007self,patel2015visual}. The limited emprical work today in transfer learning for inverse problems in imaging is promising and suggests the need for additional study. 

\paragraph{Nonlinear inverse problems.} Nearly all the examples presented in this tutorial had linear operators for the forward model. However, in many applications the true forward model is nonlinear. A number of works have investigated the phase retrieval problem~\cite{sinha2017lensless,rivenson2018phase,metzler2018prdeep,hand2018phase,jagatap2019phase,metzler2020S3PR,deng2020learning}, often with great empirical success. However, little is known about how we should solve nonlinear inverse problems generally, or the inherent tradeoffs associated with nonlinear forward models.

\paragraph{Uncertainty quantification.}
Characterizing uncertainty in solutions to an inverse problem is essential for many imaging tasks, including medical diagnosis from CT or MR images. 
However, most learning approaches investigated in this work do not provide uncertainty estimates. Taking a Bayesian perspective, recent work \cite{adler2018deep,schlemper2018bayesian} addresses this shortcoming by estimating a full posterior distribution of images fitting a given set of measurements (or estimate statistics derived from the posterior) using an generative adversarial training framework. This is used in \cite{adler2018deep} to give pixel-wise variance estimates and perform hypothesis testing in a CT reconstruction setting. Incorporating similar uncertainty quantification into other learning-based approaches, especially in cases where less is known about the distribution of ground truth images, is an interesting open problem.

\subsection*{Acknowledgments} We thank Davis Gilton for performing the MRI reconstruction experiments pictured in Figures \ref{fig:robust_forward} and \ref{fig:stability} and for his helpful feedback.
R.W.\ and G.O.\ were supported in part by  AFOSR FA9550‐18‐1‐0166,
DOE DE‐AC02‐06CH11357,
NSF OAC‐1934637 and NSF
DMS‐1930049.
C.M.\ was supported by an ORISE Intelligence Community Postdoctoral Fellowship.
A.G.D.\ and A.J.\ were supported in part by NSF CCF-1618689, DMS-1723052, CCF 1763702, 
AF 1901292, Western Digital and the Fluor Centennial Teaching Fellowship. 
R.B.\ was supported in part by NSF CCF-1911094, IIS-1838177, IIS-1730574; 
ONR N00014-18-12571 and N00014-17-1-2551;
AFOSR FA9550-18-1-0478; and a Vannevar Bush Faculty Fellowship, ONR N00014-18-1-2047.

\bibliographystyle{IEEEtran}
\bibliography{IEEEabrv,refs}

\end{document}

%% file: macros-new.tex
\def\eps{\epsilon}
\def\reals{\mathbb{R}}

\def\bx{\bm{x}}
\def\bxhat{\hat{\bx}}
\def\by{\bm{y}}
\def\bz{\bm{z}}

\def\cA{\mathcal{A}}

\def\T{\top}

\def\ie{{\em i.e.,~}}
\def\eg{{\em e.g.,~}}
\def\bxstar{\bx^\star}

\newcommand{\vareps}{\varepsilon}

\DeclareMathOperator*{\argmin}{arg\,min}

\newenvironment{squishlist}
{   \begin{list}{$\bullet$}
    { \setlength{\itemsep}{0pt}      \setlength{\parsep}{2pt}
      \setlength{\topsep}{0pt}       \setlength{\partopsep}{0pt}
      \setlength{\leftmargin}{1.5em} \setlength{\labelwidth}{1em}
      \setlength{\labelsep}{0.5em} } }
      {\end{list}}
      
\newcommand{\secref}[1]{\S\ref{#1}}

%% file: applications.tex
\rowcolors{2}{gray!40}{white}

\begin{table}[ht!]
    \caption{\small\sl Examples of inverse problems in imaging}
    \label{tab:applications}
    \centering
\begin{tabular}{p{1.25in}|p{1.5in}|p{3.2in}}
\rowcolor{midnight}
  \bem{white}{\bf{Application}} & \bem{white}{\bf{Forward model}} &  \bem{white}{\bf{Notes}} \\ 
Denoising \cite{gonzalez2007image} & $A=I$ &$I$ is the identity matrix\\
Deconvolution \cite{gonzalez2007image,caroli1987coded} & $\cA(\bx) = \bm{h}* \bx$ & $\bm h$ is a known blur kernel and $*$ denotes convolution. When $\bm h$ is unknown the reconstruction problem is known as blind deconvolution.\\
Superresolution \cite{farsiu2004fast,yang2010image} & $A = SB$& $S$ is a subsampling operator (identity matrix with missing rows) and $B$ is a blurring operator cooresponding to convolution with a blur kernel \\
Inpainting \cite{bertalmio2000image} & $A = S$ & $S$ is a diagonal matrix where $S_{i,i}=1$ for the pixels that are sampled and $S_{i,i}=0$ for the pixels that are not.\\
Compressive Sensing \cite{candes2006stable,duarte2008single} & $A = SF$ or $A = $ Gaussian or Bernoulli ensemble & $S$ is a subsampling operator (identity matrix with missing rows) and $F$ discrete Fourier transform matrix.\\
MRI \cite{fessler2010model}  & $A = SFD$ & $S$ is a subsampling operator (identity matrix with missing rows), $F$ is the discrete Fourier transform matrix, and $D$ is a diagonal matrix representing a spatial domain multiplication with the coil sensitivity map (assuming a single coil aquisition with Cartesian sampling in a SENSE framework \cite{pruessmann1999sense}).\\
Computed tomography \cite{gonzalez2007image} & $A = R$ & $R$ is the discrete Radon transform \cite{beylkin1987discrete}.
\\
Phase Retrieval \cite{fienup1987phase,bertolotti2012non,tian2014multiplexed,metzler2017coherent} & $\cA(\bx) = |A\bx|^2$ & $|\cdot|$ denotes the absolute value, the square is taken elementwise, and $A$ is a (potentially complex-valued) measurement matrix that depends on the application. 
The measurement matrix ${A}$ is often a variation on a discrete Fourier transform matrix.\\
\end{tabular}
\end{table}

%% file: taxonomy_table.tex
\rowcolors{2}{gray!40}{white}

\begin{table}[ht!]

    \caption{\small\sl Major categories of methods learning to solve inverse problems based on what is known about the forward model $\cA$ and the nature of the training data, with examples for each. Details are described throughout Section~\ref{sec:taxonomy}.
    }
    \label{tab:taxonomy}
    \centering
    \footnotesize

\begin{tabular}{p{1.25in}|p{1.1in}|p{1.1in}|p{1.1in}|p{1.1in}}
\rowcolor{midnight}
&  \bem{white}
{\small \bf{Supervised with matched $(\bx,\by)$ pairs}} &
\bem{white}
{\bf{\small Train from unpaired $\bx$'s and $\by$'s  (Unpaired ground truths and Measurements)}} 
&\bem{white}
{\bf{\small Train from $\bx$'s only (Ground truth only)}} &  \bem{white}
{\bf{\small Train from $\by$'s only (Measurements only)}}  \\
{\cellcolor{midnight!70}{\bem{white}
{\bf \small $\cA$ fully known during training and testing (\secref{sec:Aknown})} }} &
\bem{midnight}{\secref{sec:1a}:} Denoising auto-encoders~\cite{mousavi2015deep}, 
U-Net~\cite{jin2017deep}, Deep convolutional framelets~\cite{ye2018deep}
Unrolled optimization~\cite{diamond2017unrolled,meinhardt2017learning,aggarwal2018modl,adler2018learned}, Neumann networks \cite{gilton2019neumann}
& {\em amounts to training from $(\bx,\by)$ pairs}& {\em amounts to training from $(\bx,\by)$ pairs}&
\bem{midnight}{\secref{sec:1d}:}
SURE LDAMP \cite{metzler2018unsupervised,zhussip2019training},  Deep Basis Pursuit~\cite{tamir2019unsupervised}\\
{\cellcolor{midnight}{\bem{white}
{\bf \small $\cA$ known only at test time (\secref{sec:Atest})}}} & \bem{midnight}{\secref{sec:2abd}} & \bem{midnight}{\secref{sec:2abd}}& \bem{midnight}{\secref{sec:2c}:} CSGM~\cite{bora2017compressed}, LDAMP~\cite{metzler2017learned}, OneNet~\cite{rick2017one}, Plug-and-play \cite{venkatakrishnan2013plug}, RED \cite{romano2017little} &   \bem{midnight}{\secref{sec:2abd}} \\
{\cellcolor{midnight!70}{\bem{white}
{\bf \small $\cA$ partially known (\secref{sec:Apartial})}}} & \bem{midnight}{\secref{sec:3a}} & \bem{midnight}{\secref{sec:3b}:} 
CycleGAN \cite{engin2018cycle}
&  \bem{midnight}{\secref{sec:3c}:}  Blind deconvolution with GAN's \cite{kupyn2018deblurgan,asim2018blind,anirudh2018unsupervised} & \bem{midnight}{\secref{sec:3d}:} AmbientGAN\cite{bora2018ambientgan},
Noise2Noise \cite{lehtinen2018noise2noise}, UAIR \cite{pajot2018unsupervised}\\
{\cellcolor{midnight}{\bem{white}
{\bf \small $\cA$ unknown (\secref{sec:Aunknown})}}} & \bem{midnight}{\secref{sec:4a}:} AUTOMAP \cite{zhu2018image} & \bem{midnight}{\secref{sec:4bcd}}  & \bem{midnight}{\secref{sec:4bcd}}  &  \bem{midnight}{\secref{sec:4bcd}}  \\
\end{tabular}
\end{table}